\documentclass[fleqn,usenatbib]{mnras}

\usepackage{newtxtext,newtxmath}

\usepackage[T1]{fontenc}

\DeclareRobustCommand{\VAN}[3]{#2}
\let\VANthebibliography\thebibliography
\def\thebibliography{\DeclareRobustCommand{\VAN}[3]{##3}\VANthebibliography}

\usepackage{graphicx}	
\usepackage{ulem}
\usepackage{amsmath}	
\usepackage[dvipsnames]{xcolor}
\usepackage{tablefootnote}
\usepackage{threeparttable}

\newcommand{\OII}{\hbox{{\rm O}{\sc \,ii}}}

\newcommand{\OIII}{\hbox{{\rm O}{\sc \,iii}}}

\newcommand{\MgII}{\hbox{{\rm Mg}{\sc \,ii}}}
\newcommand{\CIV}{\hbox{{\rm C}{\sc \,iv}}}

\newcommand{\HI}{\hbox{{\rm H}{\sc \,i}}}

\newcommand{\Ly}{\hbox{{\rm Ly}$\alpha$}}

\newcommand{\Hb}{\hbox{{\rm H}$\beta$}}

\newcommand{\kms}{\hbox{km~s$^{-1}$}}

\newcommand{\REW}{\hbox{$W_{\rm r}^{2796}$}}

\title[\MgII\ absorption in dense environments]{MusE GAs FLOw and Wind (MEGAFLOW) X. The cool gas and covering fraction of \MgII\ in galaxy groups.}

\author[M. Cherrey et al.]{
Maxime Cherrey$^{1}$\thanks{E-mail: maxime.cherrey@univ-lyon1.fr},
Nicolas F. Bouché$^{1}$,
Johannes Zabl$^{1,2}$,
Ilane Schroetter$^{3}$,
Martin Wendt$^{4,5}$,
Ivanna Langan$^{6}$,
\newauthor
Johan Richard$^{1}$,
Joop Schaye$^{7}$,
Wilfried Mercier$^{3}$,
Benoît Epinat$^{8,9}$,
Thierry Contini$^{3}$.
\\
$^{1}$Univ of Lyon1, Ens de Lyon, CNRS, Centre de Recherche Astrophysique de Lyon (CRAL) UMR5574, F-69230 Saint- Genis-Laval, France\\
$^{2}$Institute for Computational Astrophysics and Department of Astronomy \& Physics, Saint Mary’s University, 923 Robie Street, Halifax, Nova Scotia, B3H 3C3, Canada \\
$^{3}$Institut de Recherche en Astrophysique et Plan\'etologie (IRAP), Universit\'e de Toulouse, CNRS, UPS, F-31400 Toulouse, France\\
$^{4}$Institut f\"{u}r Physik und Astronomie, Universit\"{a}t Potsdam, Karl-Liebknecht-Str. 24/25, 14476 Golm, Germany \\
$^{5}$Leibniz-Institut f\"{u}r Astrophysik Potsdam, An der Sternwarte 16, D-14482 Potsdam, Germany\\
$^{6}$European Southern Observatory, Karl-Schwarzschild-Str. 2, D-85748, Garching, Germany\\
$^{7}$Leiden Observatory, Leiden University, P.O.Box 9513, NL-2300 AA Leiden, The Netherlands\\
$^{8}$Aix Marseille Univ., CNRS, CNES, LAM, Marseille, France\\
$^{9}$Canada-France-Hawaii Telescope, CNRS, 96743 Kamuela, HI, USA\\
}

\date{Accepted XXX. Received YYY; in original form ZZZ}

\pubyear{2023}

\begin{document}
\label{firstpage}
\pagerange{\pageref{firstpage}--\pageref{lastpage}}
\maketitle

\begin{abstract}

We present a study of the cool gas ($\approx 10^4$ K)  traced by \MgII\ absorptions around groups of galaxies in the MEGAFLOW  survey. Using a combination of two algorithms we blindly identify 32 groups of more than 5 galaxies at $0.3 < z < 1.5$ with $10.7 < \log_{10}(M/\rm M_{\odot}) < 13.7$. Among them 26 can be used to study potential counterpart \MgII\ absorptions. We report that 21 out of the total 120 \MgII\ absorption systems present in MEGAFLOW are associated with groups. We observe that the \MgII\ rest-frame equivalent width (\REW) drops at an impact parameter of $\approx 150$ projected kpc from the closest galaxy and $\approx$ one virial radius from the identified group center indicating that \MgII\ halos scale with the mass of the groups. 
The impact parameter where the covering fraction exceeds $50\%$   is $\log_{10}(b/\rm kpc) = 2.17 \pm 0.47$ $(2 \sigma)$ and $(b/R_{\rm vir}) = 1.67 \pm 0.98$,
which is $\approx 3$ times larger than for field galaxies ($\log_{10}(b/\rm kpc)=1.67\pm0.15$). Finally, we estimate the cool gas column density profile in groups
(from the \REW) and show that its shape follows closely the typical dark matter column density profile for halos at similar redshift and masses.
\end{abstract}



\section{Introduction}
\label{sec:introduction}

The detection of the \MgII\ $\lambda \lambda [2796, 2803]$ absorption doublet in the spectra of background quasars is one of the most efficient way to study the cool diffuse gas surrounding foreground galaxies or groups of galaxies. Indeed the low ionization potential  of the magnesium (7.6 eV) makes it a good tracer of the cool photo-ionized gas at $T\approx{10^4}$~K and hence of \HI\ \citep{Chelouche_2009, Fukugita_2017} that constitutes the major part of the mass of the Circumgalactic Medium (CGM). The \MgII\ doublet has the advantage to be detectable in the optical from the ground at intermediate redshifts $0.3 \lesssim z \lesssim 1.8$.

\MgII\ absorption systems have played a crucial role in revealing the an-isotropic nature of the CGM, representing accretion along the galactic plane and bi-conical outflows \citep{Bordoloi_2011, Bouche_2012, Kacprzak_2012, Tumlinson_2017, Zabl_2019, Schroetter_2019, Zabl_2021}.

However galaxies are not isolated objects. they are naturally clustered due to the hierarchical formation of large scale structures. A number of them live in groups ($\lesssim 50$ members) or clusters located at the nodes of the cosmic web and it is still not clear if \MgII\ absorption systems are mainly associated with these over-dense regions. Indeed, even if several works revealed that \MgII\ absorptions are often associated with multiple galaxies \citep{Nielsen_2018, Dutta_2020, Hamanowicz_2020}, one can wonder if these observations can be explained by the natural correlation function or if they probe a favored presence of absorptions around over-densities.

A closely related question is what drives the strength of absorptions in these environments: the galaxy properties or the dark matter halos? Pioneering works \citep{Churchill_1999a, Charlton_2000, Rigby_2002} revealed the prevalence of weak \MgII\ absorbers and the diversity of absorption systems, that are often composed of several \MgII\ clouds. Later, \citet{Bouche_2006} followed by \citet{Lundgren_2009} and \citet{Gauthier_2009} showed from purely statistical considerations that the halo mass is anti-correlated with the \MgII\ absorption rest-frame equivalent width (\REW). This result indicates that strong \MgII\ systems are not virialized but preferentially produced by outflows associated with individual galaxies rather than big halos. This picture was reinforced by several observations of strong absorptions probably caused by outflows from individual galaxies \citep{Nestor_2011, Guha_2022}. In group environments, the absorption strength would hence arise from the added contributions of the individual galaxies \citep{Bordoloi_2011, Fossati_2019}. However the study of the absorption kinematics in recent works points toward a more complex situation \citep{Nielsen_2018}. Indeed several individual cases  \citep{Kacprzak_2010_ultra, Gauthier_2013, Bielby_2017, Epinat_2018, Leclercq_2022, Nielsen_2022} revealed a complex intragroup medium affected both by outflows and various interactions. 
Furthermore, for more massive structures like clusters, the strength of the \MgII{} absorption seems not to be correlated with their mass \citep{Mishra_2022} nor the star formation rate (SFR) of the closest neighbour \citep{Anand_2022} and would thus be rather caused by interactions or intracluster media.

It is important to disentangle the strength (column density and kinematics), the probability and the spatial extent of the absorptions. Several works clearly found an anti-correlation of \MgII\ absorption strength versus impact parameter for isolated galaxies or field galaxies but not for groups \citep{Chen_2010, Nielsen_2018, Huang_2021} indicating that the \MgII\ halos would extend further in these environments \citep{Bordoloi_2011}. Recent works also revealed that the probability to have an absorption associated with a group is significantly higher than for isolated galaxies \citep{Nielsen_2018, Dutta_2020,Dutta_2021} at similar impact parameter.

The above conclusions on \MgII\ absorptions in dense environments are often difficult to draw for two main reasons. First the definition of what is a group is not always the same and in many cases it simply consists in having two or more galaxies in the field of view of the instrument (which implies that the definition depends on the field of view). Second because many surveys are absorption-centric, meaning that the groups/galaxies counterparts are only searched in the vicinity of the known absorptions. 

We propose here to study the cool gas around groups in the {MusE GAs FLOw and Wind} survey (MEGAFLOW, desribed in Section \ref{sec:megaflow}) with an approach that remedies to these two issues. For that we first quantify clearly what is an over-density by using the two point correlation function and identify blindly all the groups in MEGAFLOW using a combination of two algorithms (Section \ref{sec:group_identification}). We then study potential \MgII\ absorption counterparts (Section \ref{sec:mgii_absorption}) and look at the \MgII\ absorption profile. From that we estimate the \HI\ column density profile and compare it to the dark matter column density profile for a halo of similar mass (Section \ref{sec:HI_column}). Finally we compute the \MgII{} covering fraction around groups (Section \ref{sec:covering_fraction}) and compare our results to the existing literature (Section \ref{sec:discussion}). Our conclusions are presented in Section \ref{sec:conclusion}.

This approach is made possible by using VLT/MUSE as it offers the possibility to identify all galaxies down to the detection limit around a quasar LOS by scanning spectral cubes within a field of view of $1\times1$ arcmin\textsuperscript{2} in the 4700 - 9350~\AA\ wavelength range. The \MgII\ absorption lines are detected in the quasars spectra using high resolution spectroscopy performed with UVES \citep{Dekker_2000} in the range 3000 - 11000~\AA.

In all this article, we use a standard flat $\Lambda$CDM cosmology with $H_0 = 69.3$~km~s\textsuperscript{-1} Mpc\textsuperscript{-1}, $\Omega_M = 0.29$, $\Omega_{\Lambda} = 0.71$  \citep[see][]{WMAP_2013}, and the distances are all given in proper kpc.\\

\section{The MEGAFLOW survey}
\label{sec:megaflow}
The present work is based on the MEGAFLOW survey \citep[][Bouche et al. 2023 in prep]{Schroetter_2016, Zabl_2019, Schroetter_2019}, that aims at building a large \MgII\ absorptions - galaxies sample using combined observations from VLT/MUSE and VLT/UVES in 22 quasar fields. These quasars were identified in the \citet{Zhu_Menard_2013} catalog built with SDSS spectral observations. They were selected because they have multiple ($\geq 3$) strong \MgII\ absorptions (\REW\ $>0.5$ \AA) at redshifts $0.3 < z < 1.5$ such that the corresponding [\OII] doublet of their galaxy counterparts fall in the $4700 - 9350$ \AA{}  range of MUSE. These selected quasars finally represent a total of 79 strong \MgII\ absorption systems that constitute the MEGAFLOW DR1 catalog.

Follow-up observations were performed between 2014 and 2016 for each quasar using the VLT/UVES echelle spectrograph in order to obtain high-resolution (R $\approx{38000}$, pixel size $\approx 1.3~\kms$) 1D spectra. These observations were used to identify systematically all the \MgII\ absorption systems in the 22 fields down to a detection limit of \REW\ $\approx 0.1$ \AA~. Finally, 48 new absorption systems have been detected and added to the 79 already known strong absorptions to form a total of 127 absorptions that constitute the MEGAFLOW DR2 catalog. Among them 120 have low redshifts  $z<1.5$. For each absorption system, \REW\ was estimated with the evolutionary algorithm from \citet{quast_2005} that models each absorption component as a Gaussian.

MUSE observations were performed between September 2014 and May 2017 during the Guaranteed Time of Observation (GTO) and using the Wide Field Mode. Adaptive Optics were used for 13 of the 22 fields. The cumulated exposure time per field ranges from 1h40 to 11h. The data reduction was performed using the ESO MUSE pipeline v1.6 \citep{Weilbacher_2012, Weilbacher_2014, Weilbacher_2016} and is described in detail in \citet{Schroetter_2016}, \citet{Zabl_2019} and Bouche et al. 2023 in prep.

In total, 2460 galaxies have been detected in the 22 quasar fields using both white light images and narrow band images produced by an algorithm that detects emission and absorption lines such as [\OII], \Hb, Ca H\&K, \Ly\ and/or [\OIII] \citep[for a detailed description of the source detection process see][]{Zabl_2019}. The redshift of the galaxies have been estimated by fitting their emission lines with a typical precision better than $\approx 30~\kms$ at $z \approx 1$. Thanks to this double detection process the MEGAFLOW sample is not biased against either passive or star-forming galaxies and is $50\%$ complete to r-mag $\approx 25.5$ and to $7.7 \times 10^{-18}$ erg s$^{-1}$ cm$^{-2}$ for [\OII] \citep[Bouche et al. 2023 in prep][]{}.

For this work, we are only interested in the 1208 galaxies that are located in the foreground of the quasars so we can study possible counterpart \MgII\ absorptions. Most of them have a redshift $0.3 < z < 1.5$ for which the [\OII] lines fall in the range of MUSE. The [\OII] flux detection limit corresponds to an un-obscured SFR limit of $0.07 ~\text{M}_{\odot}~\text{yr}^{-1}$. The stellar masses of the galaxies are estimated, when possible, using the SED fitting algorithm coniecto \citep[for details see][]{Zabl_2016} based on the stellar continuum and assuming a Chabrier Initial Mass Function \citep[][]{Chabrier_2003}. The estimated stellar masses in MEGAFLOW range from $10^{6}~\text{M}_{\odot}$ to $10^{12}~\text{M}_{\odot}$ with a mean at $10^{9.3}~\text{M}_{\odot}$.\\

\section{Group identification}
\label{sec:group_identification}

\subsection{Characterization of over-densities}
\label{sec:overdensities}

One of the difficulties while studying dense environments is to identify and to quantify local over-densities in the first place.
A common way to proceed is to count the number of galaxies in the Field of View (FOV) around a given redshift. If this number is above a given threshold, then these galaxies are  declared to belong to a group/an over-density. However, the threshold value is highly dependent on the size of the FOV of the instrument and must be chosen carefully to take into account the natural clustering present for all types of galaxies, even in non over-dense regions.

In order to   quantify the number of galaxies that we expect in the MUSE FOV, we use the two-point correlation function $\xi(r)$ which, by definition, gives the excess~\footnote{The excess is relative to a hypothetical sample of un-clustered galaxies, i.e. distributed uniformly with $\xi(r)=0$.} probability $P$ to find a second galaxy in a volume d$V_2$ at a distance $r$ from a known galaxy position
\citep{Peebles_1980}:
\begin{equation}
    P(2|1) = \overline{n}(1+\xi(r))dV_2,
	\label{eq:dP}
\end{equation} 
where $\overline n$ is the mean number density if galaxies were not clustered. The correlation function $\xi(r)$ can be approximated by a power-law on large scales up to tens of Mpc:
\begin{equation}
    \xi(r) = \left(\frac{r}{r_0}\right)^{-\gamma}, 
	\label{eq:2pcorr}
\end{equation} 
where the slope $\gamma$ is estimated to be $\gamma\approx{1.8}$ \citep{Marulli_2013} and $r_0$ is the correlation length. The latter is directly related to the mass of the halo considered \citep[e.g.][]{Mo_White_2002}, and a large body of literature have measured  $r_0$ for a variety of galaxies and redshifts. For instance, according to \citet{Cochrane_2018}, for star-forming galaxies at $z\approx 1$ (similar to our survey), the $r_0$ value corresponding to halos of mass $M_{\rm h} = 10^{11}$~M$_{\odot}$ is measured to be $r_0\approx 3$~Mpc. On the other hand, for groups with halos of mass $M_{\rm h}=10^{13}$~M$_{\odot}$, $r_0$ is approximately $7$~Mpc .

 Using Eqs.(1)-(2), We can then compute how many galaxies above a given mass $M$ we can expect to find in a cylinder of radius $R$ and in a redshift interval  $\pm \Delta z$ around the redshift $z_0$ of a halo (this redshift interval corresponds to a distance $R_z = c \Delta z/((1+z) H(z))$ along the line of sight). For that we integrate the correlation function $\xi(r)$ over the cylinder:

\begin{equation}
P(r_{\bot}<R;|r_{\rm z}|< R_z)=\int_0^{R} \int_{-R_z}^{+R_z}2\pi r_{\bot} \left[1+\left(\frac{\sqrt{r_{\bot}^2 + r_{\rm z}^2}}{r_0}\right)^{-\gamma}\right]{\rm d}r_{\bot}{\rm d}r_{\rm z}.
\label{eq:int}
\end{equation}
The number of expected galaxies above a given mass in such a cylinder is then $P(r_{\bot}<R;|r_{\rm z}|<R_z)$ times $n(M)$, the number density of halos of mass greater than $M$ \citep[here we use ][]{Tinker_2008}. 

If we assume that a \MgII\ absorption system is associated with a halo of mass $\sim 10^{11}$~M$_\odot$ (here we do not consider an over-dense region), then we can estimate the number of galaxies that we can expect around it in the MUSE FOV. For that we can take $R$ such that the cylinder has the same area on the sky as the MUSE FOV $\approx{} 3600$~arcsec\textsuperscript{2} ($R\approx 280$~kpc at $z=1$) and $\Delta z$ corresponding to a velocity difference of $500$~\kms.

\begin{figure}
	\includegraphics[width=0.8\columnwidth]{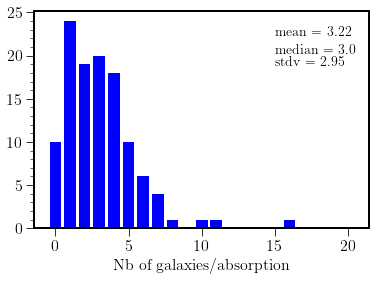}
    \caption{Distribution of the number of counterpart galaxies observed in the MUSE FOV around each MgII absorption system detected in the UVES spectra in the range $0.3 < z < 1.5$.}
    \label{fig:N_per_abs}
\end{figure}

\citet{Adelberger_2003} computed analytically the integrals in Equation \ref{eq:int} (their Equation. C2). Using their result we find that $3.3\pm{3.1}$ galaxies are expected in the MUSE FOV around an absorber in a region of mean density. It corresponds to an excess density of 14 compared to a pure random situation. The number of galaxies expected around an absorption system is presented in Table \ref{tab:r0}. We compare these values with Figure \ref{fig:N_per_abs} that shows the observed distribution of the number of galaxies within $\pm 500~\kms$ around each absorption system located at $0.3 < z < 1.5$ in MEGAFLOW. We find on average $3.2\pm{3.0}$ galaxies per absorption system in the FOV which is consistent with the expected number computed above. We also observe that it is common to have up to four galaxies around an absorption system, but the histogram then falls at five galaxies due to the MUSE FOV. Thus we consider that this value defines over-densities (i.e. not consistent with the correlation function within the MUSE FOV). In this work we aim to study the cool gas in over-dense environment so we select groups made of at least five galaxies.\\

 \begin{table}
     \centering
      \caption{Number of galaxies expected and number of galaxies found in MEGAFLOW in cylinders of radius $R$ and depth $2|\Delta v|$ centered on halos of mass $M_{\rm min}$.}
     \begin{tabular}{ccccccc}
     \hline
      $R$& $|\Delta v|$ &  $M_{\rm min}$  & $r_0$ &  $N_{\rm exp}$
        & $N_{\rm found}$ \\
     (kpc)  & (\kms) &  (M$_\odot$) & (Mpc) &  & \\
     \hline 
      280 & 500 & $10^{11}$   &  3 & 3.3$\pm$3.1 & $3.2\pm3.0$\\
       100 & 500 & $10^{11}$ & 3 & 0.9$\pm$1.3 & 1.1$\pm$1.2 \\
       \hline
     \end{tabular}
     \label{tab:r0}
 \end{table}

 One can also calculate the number of groups with halo mass above a given value $M_{\rm h}$ that we expect to find in MEGAFLOW. For that we multiply the volume of the survey by $n(M_{\rm h})$.  We obtain that $8.1 \pm 2.8$ halos of mass $M_{\rm h}>10^{13}$~M$_\odot$ are expected in MEGAFLOW. With the group finding method described below we find six groups with $M_{\rm h}>10^{13}$~M$_\odot$ which is consistent with this estimation.

\subsection{Method}
\label{sec:method}
 To obtain a robust group sample, we proceed in two steps similarly to what is proposed in \citet{Rodriguez_2020}. First, we perform a classic Friends of Friends (FoF) algorithm in order to pre-select all the galaxies potentially belonging to groups. Second we refine the groups using an iterative method inspired by the halo occupation method described in \citet{Yang_2005} (see details below). 

For the first step, we use a standard FoF algorithm with the linking lengths $\Delta D = 450$~kpc and $\Delta V = 500$~\kms as recommended by \citet{Knobel_2009} to optimize completeness and purity for the detection of groups of more than five galaxies. These values are in the high range of what can be found in the literature and we use them in order not to miss any galaxy that would belong to a group. With this FoF process, 38 groups of five or more galaxies are identified in the 22 fields of the MEGAFLOW sample.

As expected, some galaxies of the groups obtained with the simple FoF algorithm are suspected to be not gravitationally bound. Indeed, in some cases, phase space-diagrams reveal groups spread over redshift ranges corresponding to velocity differences up to $1500$ km s\textsuperscript{-1} with some galaxies clearly standing out.

In order to remove the outlying galaxies, we use a process based on the halo occupation method described in \citet{Yang_2005} and later in \citet{Tinker_2021}. This process is based on the assumption, coming from both numerical simulations \citep{Jung_2022} and observations \citep{Yang_2009}, that groups are usually formed in massive Dark Matter (DM) halos often containing a massive central galaxy. The idea is then to identify the most massive galaxies as potential group centers (defined as the center of mass of the DM halo in which the group is embedded) and to compute the corresponding DM halos properties (virial mass, virial radius, virial velocity) from their stellar masses using halo mass - stellar mass relation \citep{Girelli_2020} and concentration-mass \citep{Correa_2015} relation. The nearby galaxies located in the DM halos are then considered as satellite galaxies. Based on this idea, the following algorithm is performed to refine each group previously found by the FoF method:

\begin{enumerate}
    \item If the galaxy with the highest $M_*$ has a mass larger than 1.5 times the mass of the second most massive in the group then we define it as the group center (hence the center of the halo). Otherwise we consider that there is no clear 'central galaxy' and we define the center as the group barycenter weighted by the estimated $M^*$. \label{itm:find_center}
    \item The group halo mass is estimated from the stellar mass of the most massive galaxy using the halo mass - stellar mass relation from \citet{Girelli_2020}.
    \item The probability $P_{\rm sat}$ to belong to the group is then estimated for each galaxy (see eq. \ref{eq:Psat}). 
    \item The 4 galaxies with the highest $P_{\rm sat}$ values are candidate members of the group.
    \item The halo mass of the group is recomputed from the velocity dispersion of these 5 galaxies (see eq. \ref{eq:Mvir_sigma} below). 
    \item With the new halo mass, the $P_{\rm sat}$ values are recomputed for the candidate galaxies. They are kept if $P_{\rm sat} > 0.5$.
    \item The group halo mass is updated and the $P_{\rm sat}$ values are recomputed for the remaining galaxies. The galaxy with the highest $P_{\rm sat}$ value is added to the group if this value is above 0.5.\label{itm:loop}
    \item We repeat the process from step \ref{itm:loop} to add galaxies one by one until no remaining galaxy has a $P_{\rm sat}$ value above 0.5.
\end{enumerate}

The probability $P_{\rm sat}$ to belong to the group is computed based on the DM halo properties following  \citet{Yang_2005}. In practice,
the probability $P_{\rm sat}$ to belong to the halo is computed as:

\begin{equation}
    P_{\rm sat} = 1-(1+P_{\rm proj} P_{\rm z} /B_{\rm sat})^{-1}
	\label{eq:Psat}
\end{equation}
where $B_{\rm sat}$ is the sensitivity parameter that would determine how far from the center of the halo we can go. Here we use $B_{\rm sat} \approx{} 10$ which is the value recommended by \cite{Yang_2005}. $P_{\rm proj}$ and $P_{\rm z}$ are the pseudo-probabilities corresponding to the projected and the line of sight directions respectively. 

$P_{\rm proj}$ at a given projected distance $R_p$ from the center of the halo is given by:

\begin{equation}
    P_{\rm proj}(R_p) = 2 r_s \overline{\delta} f(R_p/rs),
	\label{eq:Pproj}
\end{equation}
where $f$ is a function defined as:

\begin{equation}
    f(x) = 
    \begin{cases}
       \frac{1}{x^2-1}\left( 1- \frac{\ln{\frac{1+\sqrt{1-x^2}}{x}}}{\sqrt{1-x^2}}\right) &\quad\text{if x < 1}\\
       \frac{1}{3} &\quad\text{if x} =1 \\
       \frac{1}{x^2-1}\left( 1- \frac{\arctan{\sqrt{x^2-1}}}{\sqrt{x^2-1}}\right) &\quad\text{if x > 1}\\
     \end{cases}
	\label{eq:f}
\end{equation}
and where $\overline{\delta}$ is the over-density corresponding to an isotropic Navarro-Frenk-White \citep[NFW, see][]{NFW_1997} DM profile defined as:

\begin{equation}
    \rho_{\rm NFW}=\frac{\overline{\rho}\overline{\delta}}{\frac{r}{r_s}\left(1+\frac{r}{r_s}\right)^2},
	\label{eq:NFW}
\end{equation}
where $r_s$ is the characteristic scale parameter and $\overline{\rho}$ is the mean density of the universe.\\

$P_{\rm z}$ at a given redshift separation $\Delta z$ from the center is given by:

\begin{equation}
    P_{\rm z}(\Delta_z) = \frac{c}{\sqrt{2\pi} \sigma_v} \exp{\left[ \frac{-\Delta v^2}{2 \sigma_v^2}\right]},
	\label{eq:Pz}
\end{equation}
where $c$ is the speed of light, $\Delta v = c \Delta z / (1+z)$ is the velocity relative to the center and $\sigma_v$ is the velocity dispersion of the galaxies within the group, assumed to be $\sigma_v = V_{\rm vir}/\sqrt{2}$ with $V_{\rm vir} = (GM_{\rm vir}/R_{\rm vir})^{1/2}$.\\

The masses of the groups derived at step \ref{itm:loop} are estimated from the velocity dispersion of their members and their spatial extent. Indeed, under the assumption that a group is virialized, its mass can be related to the velocity dispersion of its galaxies along the line of sight $\sigma_{\rm los}$ and its radius $R_{\rm group}$:

\begin{equation}
    M_{\rm vir} = \frac{A R_{\rm group}\sigma_{\rm los}^2}{G},
    \label{eq:Mvir_sigma}
\end{equation}
Where $R_{\rm group}$ is estimated by taking the dispersion of the projected distance of the galaxies. The factor $A$ must be taken such that the mass estimator is unbiased. Calibration tests using groups from TNG50 lead to a choice of $A = 5.0$ which is also the value recommended by \citet{Eke_2004}.

The virial radius of the groups are derived from their virial masses:

\begin{equation}
    R_{\rm vir} = \left(\frac{3 M_{\rm vir}}{4 \pi \Delta_{\rm vir} \rho_c(z)}\right)^{1/3}
    \label{eq:Mvir_Rvir}
\end{equation}
where $\rho_c(z)$ is the critical density of the universe at redshift $z$ and $\Delta_{\rm vir} = 18 \pi^2 + 82x-39x^2$ with $x = \Omega_M(z) - 1$ applicable for a flat universe with $\Omega_k = 0$ \citep{Bryan_1998}.

The main sources of error in our estimation of $M_{\rm vir}$ are the estimation of the velocity dispersion $\sigma_{\rm los}$ and the estimation of the projected distance dispersion $R_{\rm group}$. Under the assumption of a normal distribution, the 1-$\sigma$ uncertainty associated to an unbiased standard deviation estimator of value $x$ on a sample of size $N$ is equal to $x\sqrt{1-k^2(N)}$ \citep[see][]{Markowitz_1968} where $k(N)$ is given by:

\begin{equation}
    k(N) = \sqrt{\frac{2}{N-1}}\frac{\Gamma(N/2)}{\Gamma((N-1)/2)}
\end{equation}
with $\Gamma$ the gamma function. 
The above equation is used to estimate the uncertainty of the velocity dispersion and the projected distance dispersion. As a consequence, the error on $M_{\rm vir}$ logically increases when the number of galaxies decreases. With fewer than five galaxies, the error on the virial radius is above 30\%. For this reason and the one explained in Sect. \ref{sec:overdensities}, we focus on groups of five galaxies or more in the rest of the analysis.
The uncertainties on the virial mass are propagated to the virial radius.

One of the main limitation of the method presented here is that groups can be truncated by the MUSE FOV. In such case the center of the group could be wrong and the group members badly identified. This effect is an additional source of error that we didn't take into account in this work.

\subsection{The group sample}
From the 38 groups of more than five galaxies detected by the FoF algorithm, we finally obtain 33 groups after the refinement process. One of them is at high redshift ($z = 3.55$), the others are in the range $0.3 < z  <1.5$. We find six groups with an estimated halo mass above $10^{13}$~M$_\odot$, which is in line with the expected number estimated in Section \ref{sec:overdensities}. 

Among the 33 groups, three have the same redshift as the quasar of the field (note that among our 22 quasars, only five of them are located at redshifts below 1.5 where our groups are preferentially detected using the [\OII] emission lines). We remove these three groups from the analysis because \MgII\ absorption could be affected by the position of the quasar among the group and by the galaxy hosting the quasar (one of these groups is associated with an absorption). Another group is removed because it is located at a redshift higher than the redshift of the quasar. Three other groups located at redshift where there is no UVES coverage on \MgII\ are removed from the analysis. In total seven groups are removed and we finally obtain a sample of 26 groups that we use as a basis to study \MgII\ absorption in the quasars spectra. These groups have $\log_{10}(M_{\rm vir}/\rm M_{\odot})$ ranging from 10.7 to 13.7 with a median value of 12.3 and with redshifts ranging from 0.5 to 1.4 with a median value of 1.0. 16 out of the 26 groups have a central galaxy as defined at step \ref{itm:find_center}. The centers of the other 10 groups are the barycenters weighted by the stellar masses of the galaxies.

The group sample is presented in Table \ref{tab:groups_summary}. The individual groups are detailed in Table \ref{tab:groups} and shown in Figure \ref{fig:group_catalog_3}. The number of galaxies per group as a function of their mass and redshift is shown in Figure \ref{fig:N_vs_Mvir}.  

\begin{table}
     \centering
      \caption{Summary of the groups of more than five galaxies identified in MEGAFLOW. The left column presents the whole sample. The right column presents the sample selected to study counterpart \MgII\ absorptions.}
     \begin{tabular}{ccc}
     \hline
       & all groups &  selected groups\\
    \hline
     Number of groups  & 33 &  26 \\
     Groups with \MgII\ abs. & 22 & 21\\
     log($M_{\rm vir}$) range & 10.7 - 13.7 $M_{\odot}$ & 10.7 - 13.7 $M_{\odot}$\\
     redshift range & 0.26 - 3.55 & 0.46 - 1.43\\
     \REW range & 0.08 - 3.34 \AA & 0.08 - 3.34 \AA\\
     \hline
     \end{tabular}   
     \label{tab:groups_summary}
\end{table}

\begin{table*}
    \caption{Characteristics of the groups of more than five galaxies identified in the MEGAFLOW sample. The groups are sorted by number of galaxy members identified. The columns present the group id (1), the quasar field (2), the number of members (3), the redshift (4), the angular coordinates (5 and 6), the estimated virial mass (7), the estimated virial radius in kpc (8), the \MgII\ absorption rest-frame equivalent width in \AA\ (9), the impact parameter relative to the center of the group normalized by the virial radius (10) and the impact parameter relative to the closest galaxy in kpc (11).}
    \begin{threeparttable}[b]
    \centering
    \label{tab:groups}
    \begin{tabular}{ccccccccccc}
    \hline
      ID & field id & $N_{\rm gr}$ &  $z$  & R.A &  Dec
        & $\log_{10}{(M_{\rm{vir}}/\rm{M_{\odot}})}$ & $R_{\rm{vir}}$  & $W^{2796}_r$ & $b_{\rm{center}}/R_{\rm{vir}}$ & $b_{\rm{min}}$ \\

       (1)  & (2) & (3) & (4) & (5) & (6) &(7)& (8) & (9) & (10) & (11)\\
     \hline
     1 & J1039p0714 & 21 & 0.99 & $159^{\circ}$54'40" & $7^{\circ}$14'49" & 12.9 & 309 & 0.63 & 0.96 $\pm$ 0.10 & 46\\
     2 & J0145p1056 & 21 & 0.94 & $26^{\circ}$18'23" & $10^{\circ}$56'18" & 13.2 & 400 & 0.12 \tnote{1} & 0.21 $\pm$ 0.02 & 1\\
     3 & J0014m0028 & 13 & 0.83 & $3^{\circ}$42'52" & $-0^{\circ}$28'33" & 13.7 & 631 & 2.09 & 0.35 $\pm$ 0.05 & 8\\
     4 & J1107p1021 & 11 & 0.75 & $166^{\circ}$55'44" & $10^{\circ}$21'27" & 11.9 & 161 & 2.34 & 0.13 $\pm$ 0.11 & 37\\
     5 & J1107p1021 & 10 & 0.90 & $166^{\circ}$55'10" & $10^{\circ}$21'15" & 12.2 & 191 & <0.05 & 1.38 $\pm$ 0.21 & 93\\
     6 & J0058p0111 & 9 & 0.64 & $14^{\circ}$44'02" & $1^{\circ}$11'28" & 12.4 & 247 & 3.34 & 0.17 $\pm$ 0.04 & 6\\
     7 & J1352p0614 & 8 & 0.61 & $208^{\circ}$04'19" & $6^{\circ}$14'18" & 13.0 & 411 & 0.78 & 0.27 $\pm$ 0.05 & 10\\
     8 & J1236p0725 & 7 & 1.33 & $189^{\circ}$05'51" & $7^{\circ}$25'34" & 12.5 & 193 & 0.41 & 1.04 $\pm$ 0.20 & 113\\
     9 & J1358p1145 & 7 & 1.10 & $209^{\circ}$31'56" & $11^{\circ}$46'21" & 12.3 & 186 & <0.05 & 1.55 $\pm$ 0.32 & 253\\
     10 & J1358p1145 & 7 & 1.15 & $209^{\circ}$31'57" & $11^{\circ}$45'36" & 13.1 & 303 & - \tnote{1}& 0.93 $\pm$ 0.17 & 1\\
     11 & J0015m0751 & 7 & 0.87 & $3^{\circ}$53'37" & $-7^{\circ}$51'10" & 12.0 & 170 & - \tnote{1}& 0.57 $\pm$ 0.11 & 1\\
     12 & J1509p1506 & 7 & 0.97 & $227^{\circ}$14'55" & $15^{\circ}$06'48" & 12.7 & 275 & 1.30 & 0.43 $\pm$ 0.08 & 80\\
     13 & J2152p0625 & 7 & 1.43 & $328^{\circ}$00'13" & $6^{\circ}$25'19" & 13.1 & 310 & 1.15 & 0.35 $\pm$ 0.09 & 61\\
     14 & J2137p0012 & 6 & 1.21 & $324^{\circ}$27'04" & $0^{\circ}$12'36" & 12.6 & 224 & 1.13 & 0.63 $\pm$ 0.13 & 95\\
     15 & J1352p0614 & 6 & 1.14 & $208^{\circ}$04'29" & $6^{\circ}$14'20" & 12.6 & 242 & 1.40 & 0.49 $\pm$ 0.14 & 27\\
     16 & J0838p0257 & 6 & 0.94 & $129^{\circ}$42'56" & $2^{\circ}$57'10" & 11.8 & 141 & 0.77 & 0.47 $\pm$ 0.36 & 64\\
     17 & J1425p1209 & 6 & 0.26 & $216^{\circ}$24'35" & $12^{\circ}$09'16" & 12.4 & 309 & - \tnote{2}& 0.07 $\pm$ 0.01 & 10\\
     18 & J0131p1303 & 6 & 0.84 & $22^{\circ}$54'05" & $13^{\circ}$03'38" & 12.7 & 294 & 0.14 & 0.18 $\pm$ 0.1 & 96\\
     19 & J2137p0012 & 5 & 1.04 & $324^{\circ}$26'52" & $0^{\circ}$12'07" & 13.3 & 426 & 0.87 & 0.38 $\pm$ 0.09 & 84\\
     20 & J1314p0657 & 5 & 0.99 & $198^{\circ}$31'29" & $6^{\circ}$57'24" & 11.1 & 81 & 0.91 & 0.46 $\pm$ 0.25 & 38\\
     21 & J0015m0751 & 5 & 0.63 & $3^{\circ}$54'01" & $-7^{\circ}$50'47" & 12.9 & 338 & - \tnote{2}& 0.44 $\pm$ 0.1 & 116\\
     22 & J0800p1849 & 5 & 0.61 & $120^{\circ}$01'21" & $18^{\circ}$49'21" & 12.7 & 328 & 1.02 & 0.40 $\pm$ 0.09 & 67\\
     23 & J0014m0028 & 5 & 1.36 & $3^{\circ}$43'15" & $-0^{\circ}$28'42" & 12.3 & 171 & <0.04 & 0.77 $\pm$ 0.17 & 131\\
     24 & J1107p1021 & 5 & 1.30 & $166^{\circ}$55'48" & $10^{\circ}$21'35" & 12.0 & 137 & 0.53 & 0.68 $\pm$ 0.19 & 58\\
     25 & J0131p1303 & 5 & 1.34 & $22^{\circ}$53'31" & $13^{\circ}$03'21" & 12.2 & 157 & 0.17 & 1.99 $\pm$ 0.45 & 127\\
     26 & J0838p0257 & 5 & 0.82 & $129^{\circ}$42'45" & $2^{\circ}$56'51" & 11.8 & 147 & 0.28 & 1.05 $\pm$ 0.24 & 142\\
     27 & J2137p0012 & 5 & 0.81 & $324^{\circ}$27'17" & $0^{\circ}$12'25" & 11.8 & 152 & 0.80 & 0.59 $\pm$ 0.13 & 72\\
     28 & J0014p0912 & 5 & 1.22 & $3^{\circ}$43'03" & $9^{\circ}$12'06" & 11.6 & 103 & 1.43 & 1.59 $\pm$ 1.36 & 99\\
     29 & J0800p1849 & 5 & 0.60 & $120^{\circ}$00'52" & $18^{\circ}$49'20" & 11.9 & 173 & 0.08 & 0.85 $\pm$ 0.21 & 85\\
     30 & J0015m0751 & 5 & 0.46 & $3^{\circ}$53'59" & $-7^{\circ}$50'36" & 10.7 & 76 & <0.07 & 2.35 $\pm$ 0.53 & 167\\
     31 & J1107p1021 & 5 & 3.55 & $166^{\circ}$55'55" & $10^{\circ}$21'31" & 12.8 & 136 & - \tnote{1,2}& 0.80 $\pm$ 0.18 & 101\\
     32 & J0015m0751 & 5 & 0.53 & $3^{\circ}$54'16" & $-7^{\circ}$51'18" & 12.6 & 312 & <0.07 & 0.67 $\pm$ 0.15 & 122\\
     33 & J0014m0028 & 5 & 1.40 & $3^{\circ}$43'09" & $-0^{\circ}$28'09" & 12.9 & 258 & 0.09 \tnote{2}& 0.74 $\pm$ 0.17 & 84\\
     \hline
    \end{tabular}
    \begin{tablenotes}
       \item [1] Group redshift equal or greater than the redshfit of the quasar.
       \item [2] No or bad UVES coverage.
     \end{tablenotes}
    \end{threeparttable}
\end{table*}

\begin{figure}
	\includegraphics[width=1\columnwidth]{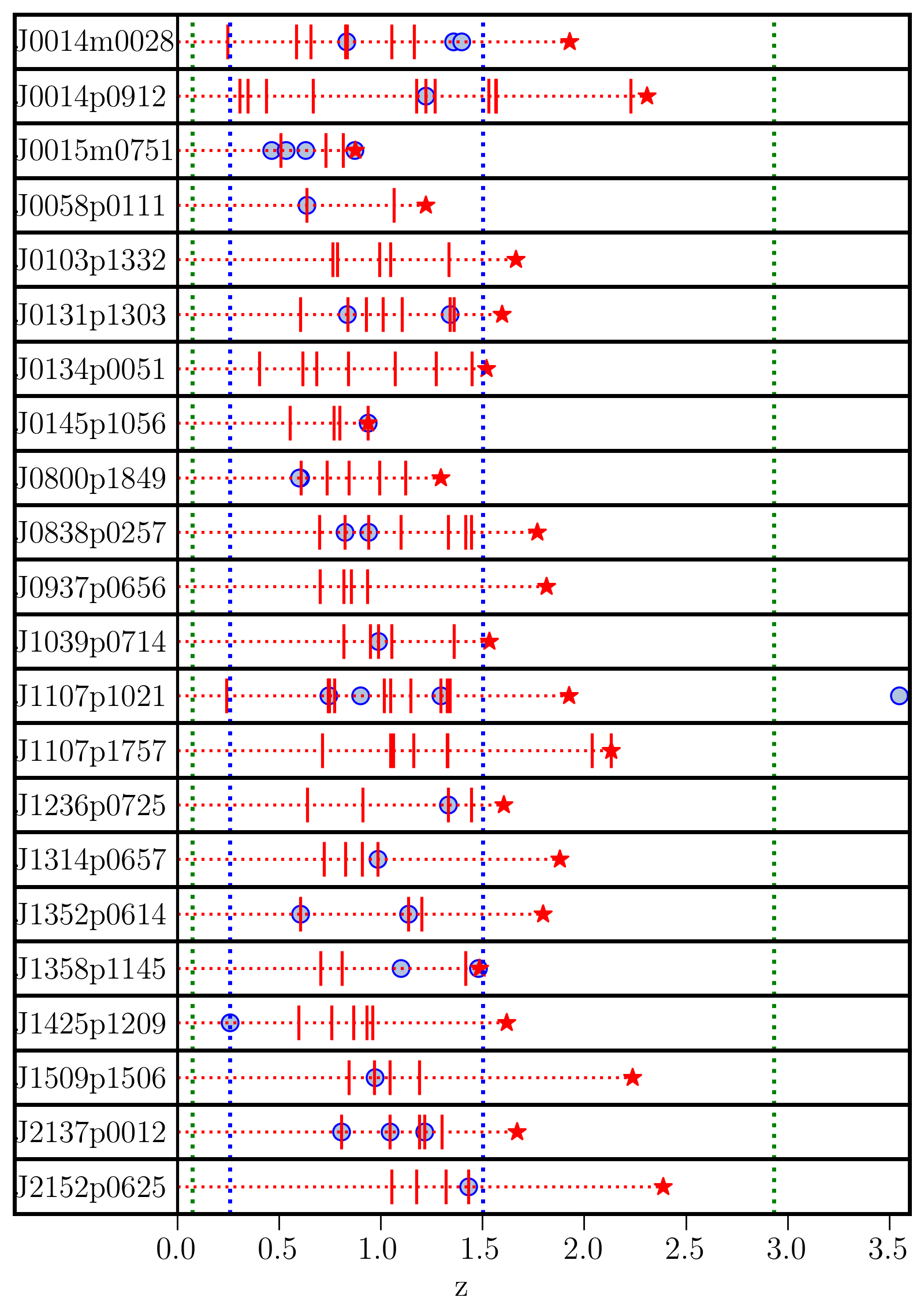}
    \caption{Groups of more than five galaxies observed in each quasar field as a function of redshift. The groups are represented by the blue circles. The quasars are represented by the red stars. The detected \MgII\ absorption systems are marked by the red vertical ticks. The blue vertical dotted lines indicate the [\OII] detection limits for MUSE. The green vertical dotted lines indicate the \MgII\ detection limits for UVES. Two groups are present at similar redshift ($\approx$ 0.61) in field J0800p1849 and cannot be distinguished on the figure.}
    \label{fig:field_description}
\end{figure}

\begin{figure}
	\includegraphics[width=0.9\columnwidth]{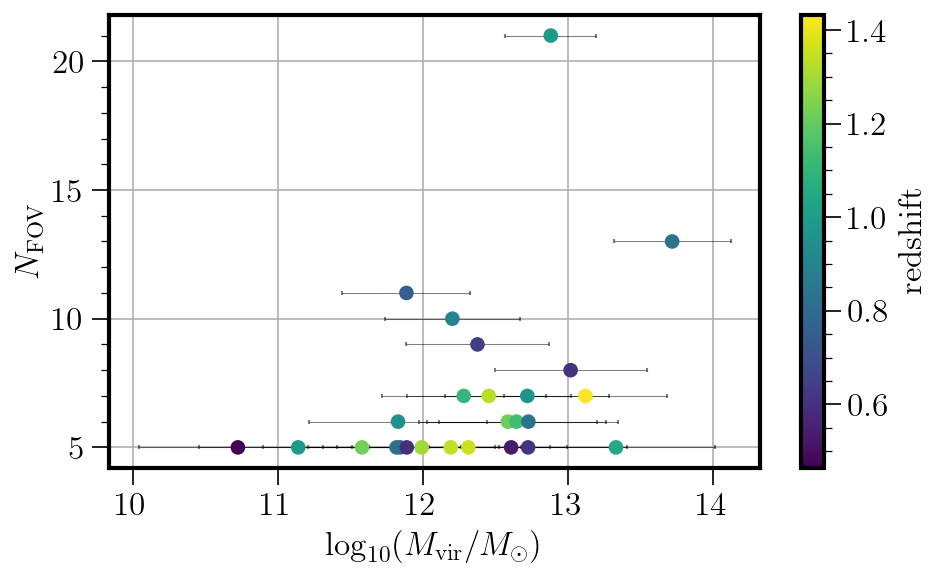}
    \caption{Number of galaxies visible in the MUSE FOV as a function of the estimated halo masses for 26 selected groups of more than five galaxies identified in MEGAFLOW. The redshift of the groups is color coded.}
    \label{fig:N_vs_Mvir}
\end{figure}

\begin{figure}
	\includegraphics[width=0.9\columnwidth]{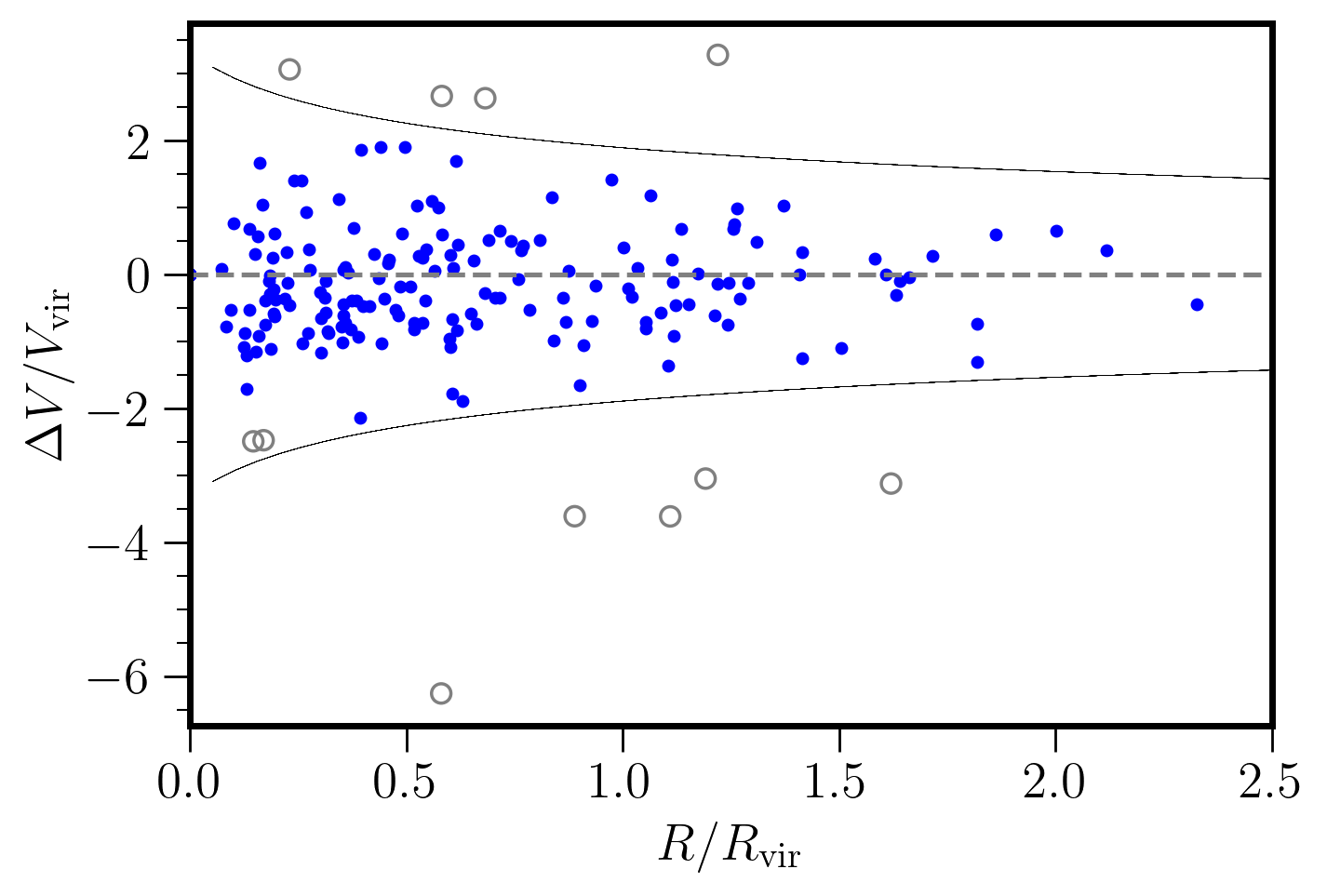}
    \caption{Superposed phase diagram of the 26 selected groups of more than five galaxies. For all the groups the galaxies are plotted in the group center rest-frame. The projected distance to the center of the group is normalized by the virial radius and the velocity difference to the center of the group is normalized by the virial velocity. The grey open circles are the nearby galaxies rejected by the algorithm. The black lines are the escape velocity caustics computed from the estimated mass of the groups assuming NFW properties.}
    \label{fig:phase_space}
\end{figure}

We can also represent each group in a phase space diagram, where each galaxy is positioned according to its projected distance and its velocity difference relative to the center of the group. The superposition of the 26 phase space diagrams is shown in Figure \ref{fig:phase_space}. We see that group members are found up to twice the virial speed and projected distances up to twice the virial radius of the groups.  

\subsubsection{Estimation of the SFR}
We estimate the Star Formation Rate (SFR) of the group members using the dust corrected relation from \citet{gilbank_2011} based on the [\OII] flux and the estimated stellar mass:

\begin{equation}
    SFR = \frac{L([\rm{OII}]_{\rm obs})/3.8\times10^{40}\rm erg/s}{a \tanh[(x-b)/c]+d}
\end{equation}
with $a = -1.424$, $b = 9.827$, $c = 0.572$, $d = 1.70$ and $x = \log(M^*)$. For 12 groups out of 33, the center corresponds to a galaxy that can be described as "passive" with a specific SFR (sSFR) below $0.1$ Gyr$^{-1}$. For 5 additional groups, the center is within 50 projected kpc from a passive galaxy. This tendency to have quenched central galaxies due to interactions or merger events is well-know \citep{tal_2014, smethurst_2017}, and tends to confirm our group center identification. The passive galaxies are indicated in red in Figure \ref{fig:group_catalog_3}.

\section{\MgII\ absorption versus impact parameter}
\label{sec:mgii_absorption}

Now that the groups have been identified in the MEGAFLOW sample, we want to study the cool gas around them by looking at \MgII\ absorption seen in nearby quasars spectra obtained with UVES. For that we consider that a group is related to a \MgII\ absorption system if the redshift difference relative to the group center is $\Delta V < 1000~\kms$. The choice of $\Delta V$ is not crucial for the analysis as long as it is large enough to capture any potential absorption in the neighborhood of the group. For our sample, the group-absorption association remains identical for $\Delta V$ ranging from $400~\kms$ to $6000~\kms$.

Out of the 26 selected groups, 21 can be paired with a \MgII\ absorption system (nine having \REW\ $> 1$\AA\ ) and five cannot be paired with any absorption system. To quantify the profile of \MgII\ halos around groups of galaxies, we want to study how \REW\ varies with the impact parameter to the LOS. However, for groups of galaxies, the notion of impact parameter is ambiguous. \citet{Bordoloi_2011} consider the impact parameter relative to the geometric group center defined as the geometric mean of the positions of the group members, or to the most massive galaxy; \citet{Nielsen_2018} consider the impact parameter relative to the closest galaxy or to the most luminous galaxy and \citet{Dutta_2020} consider the impact parameter relative to the geometric group center, in some cases normalized by the virial radius.

We can see in Figure \ref{fig:group_catalog_3} that those different definitions are not necessarily in agreement. With our approach, the groups are assumed to lie in DM halos often containing a massive central galaxy. In consequence we focus on two definitions of the impact parameter: $b_{\rm min}$, the projected distance to the closest galaxy and $b_{\rm center}$, the projected distance to the group center. Even if these two definitions are correlated, they enable us to investigate whether absorption systems are more likely affected by the CGM of individual galaxies located close to the LOS or by the presence of an intragroup medium centered on the DM halo.

Intuitively one would expect the size of the cool gas halo to be correlated with the size of the DM halo of the group (and hence with its mass). For that reason we normalize $b_{\rm center}$ by the virial radius of the group. We do not normalize $b_{\rm min}$ by the virial radius of the closest galaxy because it would require to estimate the galaxy halo mass using the $M^*-M_{\rm halo}$ relation. However the stellar mass estimate from SED fitting could be uncertain in some cases and the $M^*-M_{\rm halo}$ relation has an important scatter.  

\begin{figure}
  \includegraphics[width=8.5cm]{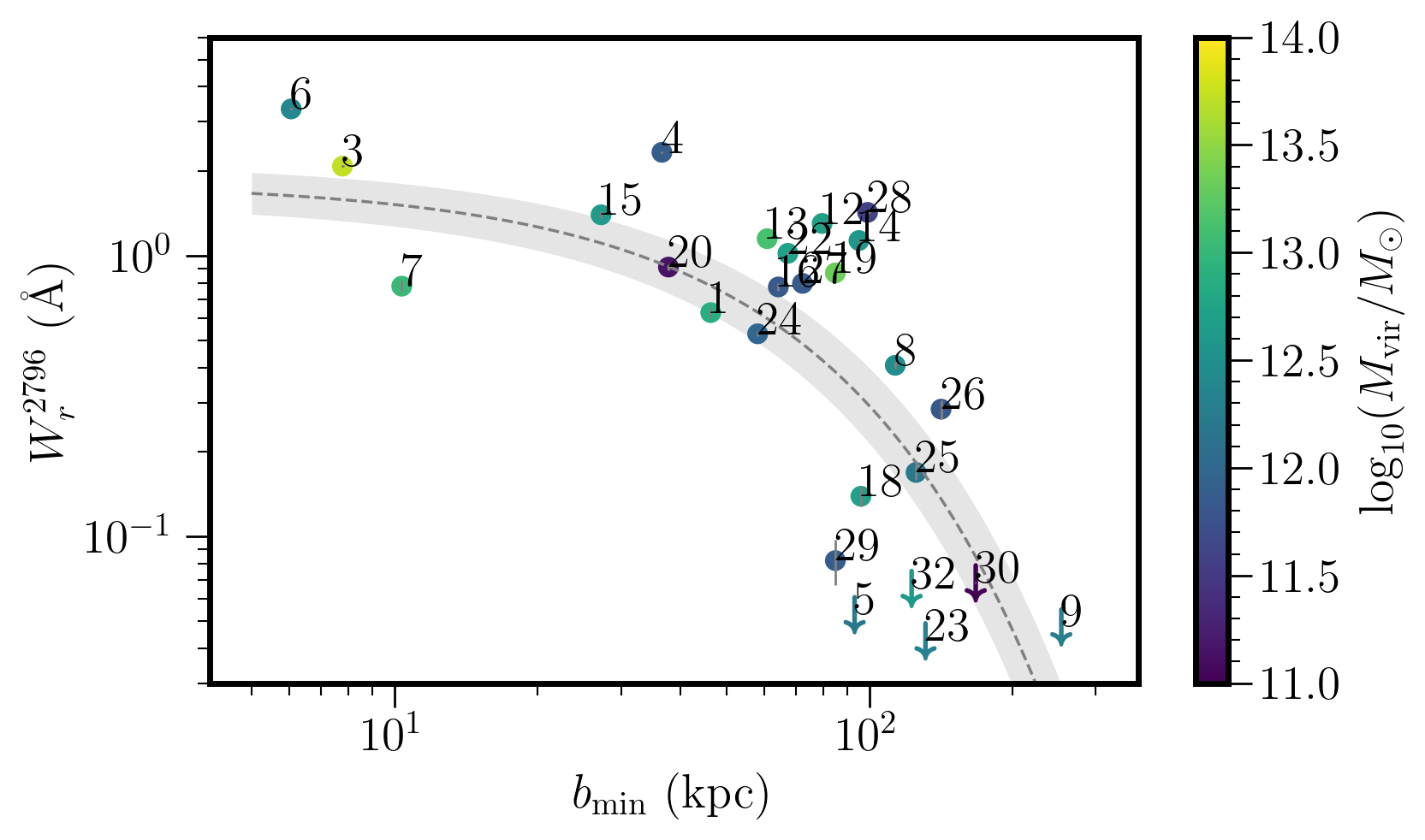}\hfill
  \includegraphics[width=8.5cm]{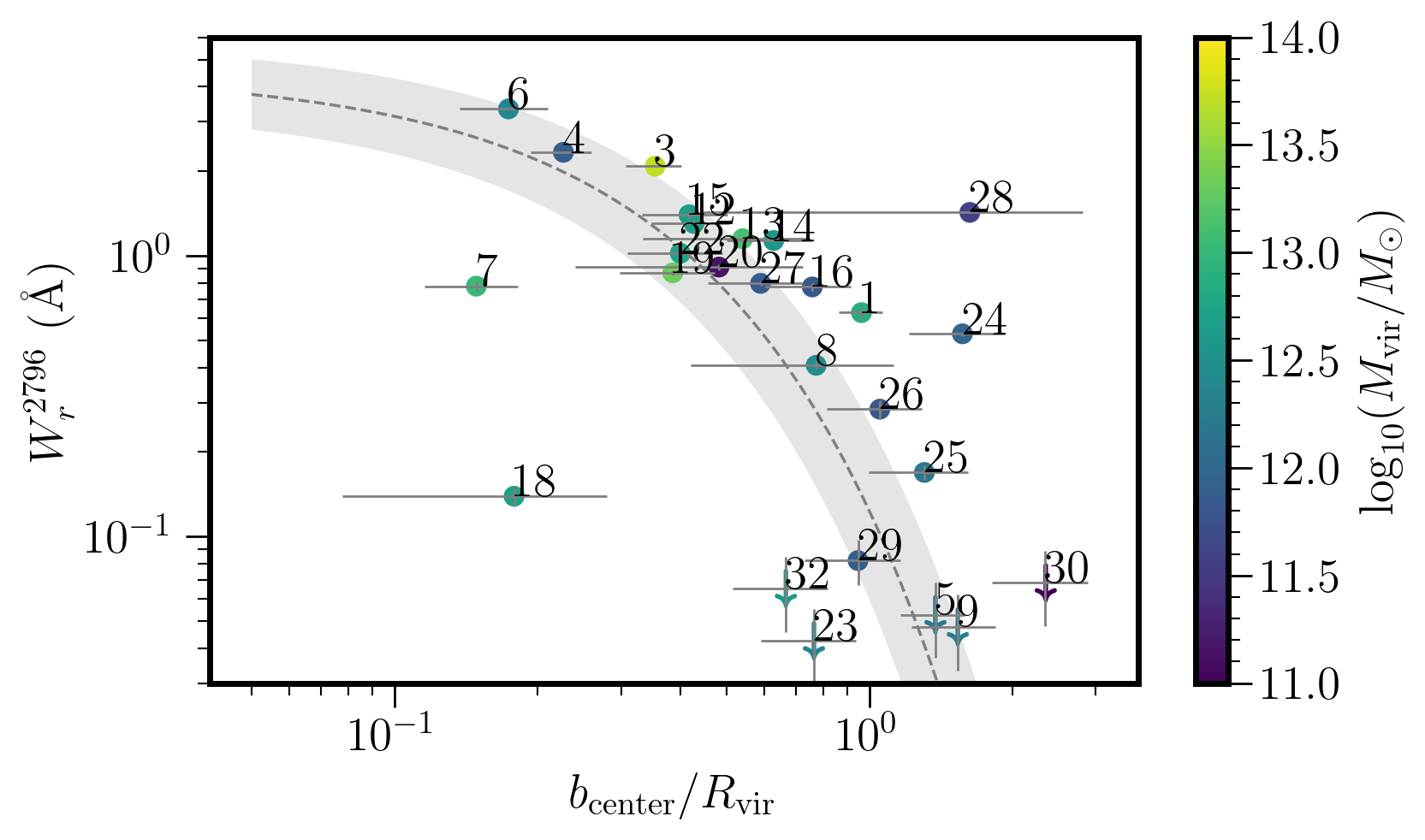}\hfill
    \caption{\MgII\ absorption rest equivalent width versus impact parameter $b$ to the closest galaxy (top) and to the group center normalized by the virial radius (bottom). The halo mass of the groups is color coded. The groups for which no \MgII\ counterpart absorption system have been detected are represented by downward arrows and plotted at the detection limit. The represented error bars are 1-$\sigma$. The grey dashed line is the best fit of the form $\log_{10}(W^{2796}_r) = A\times b + B$ and the shaded area is the corresponding 1-$\sigma$ uncertainty.}
  \label{fig:rew}
\end{figure}

\REW\ as a function of $b_{\rm min}$  and of $b_{\rm center}/R_{\rm vir}$ is shown in Figure ~\ref{fig:rew}.
The uncertainties on $b_{\rm min}$ are very small because they only consist in the precision with which the center of the quasar and the center of the closest galaxy could be determined. The uncertainties on $b_{\rm center}$ are similar but for the groups with no central galaxy identified in step \ref{itm:find_center}, they also include the propagation of the stellar mass uncertainties on the barycenter of the group. The uncertainties on $R_{\rm vir}$ are computed by propagating the uncertainties on $M_{\rm vir}$ described in Sect. \ref{sec:method} using eq. \ref{eq:Mvir_Rvir}. Figure ~\ref{fig:rew} clearly shows a scattered anti-correlation between \REW\ and impact parameter for both definitions. \REW\ seems to drop at $\approx$ 150 kpc from the closest galaxy or at the virial radius from the group center. The dispersion for the second case appears to be small even if some groups like the groups 7, 18 and 28 are standing outside of the main trend (see the discussion in section \ref{sec:discussion} for these cases).

To better characterize this decrease of \REW\ with the impact parameter, we fit it with a log-linear relation of the form:

\begin{equation}
\log{W^{2796}_r} = a + m\times b
\end{equation}
As shown in Figure~\ref{fig:rew}, some groups with low \REW\ are affected by significant vertical uncertainties due to \MgII\ absorption measurement meanwhile some groups are presenting high horizontal uncertainties when we consider $b_{\rm center}/R_{\rm vir}$. These are mostly due to poor group center or group mass estimation.
To take into account the uncertainties along the two axis, we use the results from \citet{Hogg_2010} that define the angle $\theta = \arctan(m)$ and the vector orthogonal to the linear relation 
$\hat{v}^\intercal = \begin{bmatrix}-\sin{\theta} &  \cos{\theta} \end{bmatrix}$.
A measurement $i$ of a given \MgII\ equivalent width $W^{2796}_{ri}$ (hereafter we note $W_i \doteq \log W^{2796}_{ri}$) at a given impact parameter $b_i$ can be defined by the vector $Z_i$ and the associated covariance matrix $S_i$:
\begin{equation}
Z_i = \begin{bmatrix} b_i \\ W_i \end{bmatrix},   S_i = \begin{bmatrix} \sigma_{b_i}^2 & 0 \\ 0 & \sigma_{W_i}^2. \end{bmatrix}
\end{equation}
The likelihood of such measurement can then be expressed as a function of the orthogonal displacement
$\Delta_i = \hat{v}^\intercal Z_i - a \cos{\theta}$
and of the projected covariance matrix $\Sigma_i = \hat{v}^\intercal S_i \hat{v}$. Finally, the total likelihood can be expressed as: 

\begin{equation}
\mathcal{L}(W) = K \left[ \prod_{i=1}^n  \exp{\left(\frac{\Delta(W_i)^2}{2 \Sigma_i^2}\right)} \right] \times \left[\prod_{i=1}^m \int_{-\infty}^{W_{i}}  dW' \exp{\left(\frac{\Delta(W')^2}{2 \Sigma_i^2}\right)} \right],
\end{equation}
where $K$ is a constant. The first product corresponds to the likelihood of the points that have detected \MgII\ absorption and the second products corresponds to the likelihood of the points that do not have \MgII\ absorption detected but only have an upper on $W_{i}$.

For this fit we consider that $\sigma_{W_i}$ can be decomposed into two sub-terms: a measurement uncertainty $\sigma_{m_i}$ and an intrinsic scatter $\sigma_{c}$ due to the natural variations from group to group. In consequence we express $\sigma_{W_i}$ as the quadratic sum of these two components:
\begin{equation}
    \sigma_{W_i}^2 = \sigma_{m_i}^2 + \sigma_{c}^2
\end{equation}
The intrinsic scatter $\sigma_{c}$ is estimated following \citet{Chen_2010} by comparing the deviation to the maximum likelihood solution to the measurement uncertainty:

\begin{equation}
    \sigma_c = \text{med} \left( \left[ W_i - \overline{W}(b_i) - \frac{1}{N} \sum_{j=1}^N \left(W_i - \overline{W}(b_i) \right) \right]^2 - \sigma_{m_i}^2\right).
\end{equation}
As the above equation depends on the likelihood solution, we iterate starting with $\sigma_c = 0$ until we reach convergence.

Finally, when we consider the impact parameter $b_{\rm min}$, the intrinsic scatter converges to $\sigma_c = 0.42$ dex and the best-fit parameter values are $a = 1.14 \pm{} 0.005$  and $m = -0.017 \pm{} 0.001$.

When we consider the impact parameter relative to the center of the group and normalized by the virial radius, the intrinsic scatter converges to $\sigma_c = 0.81$ dex and the best-fit parameter values are $a = 1.75 \pm{} 0.42$ and $m = -3.90 \pm{} 0.58$. For this model \REW\ drops below $0.1$ \AA\ for an impact parameter of $1.03 \times R_{\rm vir}$. The fitted models are shown along with the measured data in Figure \ref{fig:rew}.

\section{\HI\ and DM column densities}
\label{sec:HI_column}

In the previous section we have seen that the \MgII\ absorption profile seems to scale with the halo mass which is consistent with the isotherm model from \citet{Tinker_2008}. If we assume that \REW\ is proportional to the amount of cool gas along the line of sight as suggested by the works of \citet{Rao_2006} and \citet{Chelouche_2009}, it implies that the cool gas halo scales with the dark matter halo. Based on that idea we aim to compare the column density profile for these two components.

To estimate the DM column density profile we use the results from \citet{Diemer_2023}. Instead of using a standard NFW profile \citep{NFW_1997} which is not physical at high radii, they propose a functional form designed to take into account both orbiting and first in-falling DM particles as well as the asymptotic behaviour at large radii where the profile reaches the mean density of the universe. They finally suggest a form similar to a truncated Einasto profile. We use the colossus package \citep{Diemer_2018}, that implements this DM profile to compute the corresponding DM column density profile along the line of sight. For the comparison with our sample we consider a halo of mass $10^{12}$ $\text{M}_{\odot}$ at $z = 1$ (the median halo mass and redshift for our group sample are respectively $10^{12.3}$ $\text{M}_{\odot}$ and $z = 1.0$)

We then estimate the \HI\ column density from our \MgII\ absorption measurement using the results from \citet{Fukugita_2017}. They fit the correlation between \MgII\ absorption strength and \HI\ column density on a sample of \MgII\ absorptions from several catalogs with redshift $0.1<z<4.5$ for which \HI\ column densities have been measured using \HI\ absorption lines. They finally obtain the following relation:

\begin{equation}
    N_{\rm HI} = A \left(\frac{W^{2796}}{1 \text{\AA}}\right)^\alpha \left(1+z\right)^\beta,
	\label{eq:fukugita}
\end{equation}
with $\alpha = 1.69 \pm 0.13$, $\beta = 1.88 \pm 0.29$ and $A = 10^{18.96 \pm 0.10}$~cm$^{-2}$.

We use this model to estimate the \HI\ column density in our groups and we propagate the uncertainties from the relation from \citet{Fukugita_2017}. We find \HI\ column densities of approximately $10^{19}$~cm$^{-2}$ to $10^{20}$~cm$^{-2}$ for the groups where we have \MgII\ absorption detected. Our detection limit of $\approx 0.1$ Å corresponds to a \HI\ column density of approximately $2\times 10^{17}$~cm$^{-2}$. We fit the \HI\ column density profile with the method applied in section \ref{sec:mgii_absorption} on \MgII\ . For \HI\ we obtain the following parameters: $a = -14.0 \pm 0.3$, $m = -6.6 \pm 0.2$ . Figure \ref{fig:rew} shows the DM column density profile along with the \HI\ best fit and the \HI\ column densities for each group. As we can see the \HI\ and DM profiles present a very similar shape with a clear drop at the virial radius.
\\

\begin{figure}
	\includegraphics[width=8.5cm]{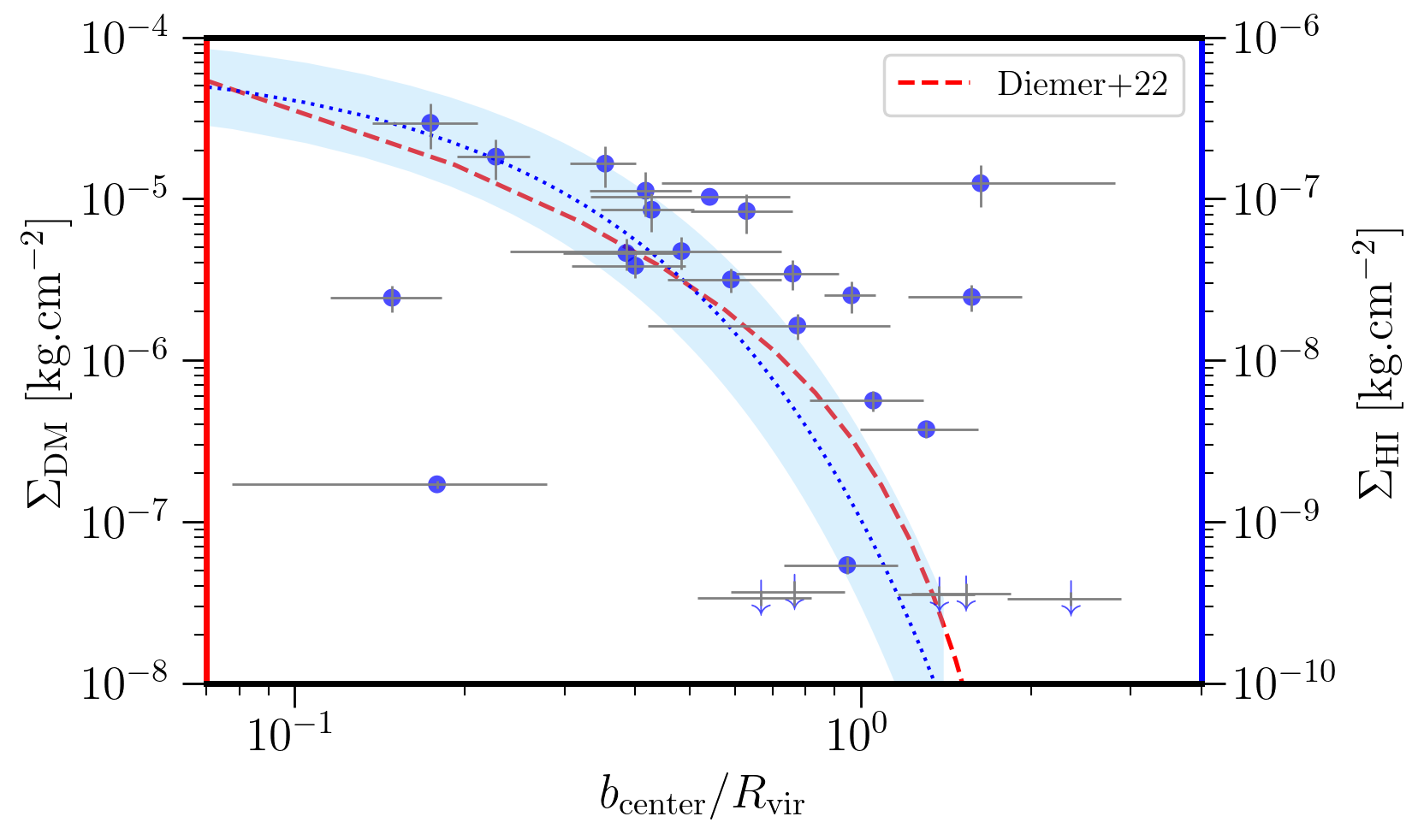}
    \caption{Dots with right axis: \HI\ column density derived from $W^{2796}_r$ based on \citet{Fukugita_2017} for the 26 groups of more than 5 galaxies. The represented error bars are 1-$\sigma$ uncertainties. The blue dotted line is
    the best fit of the form $\log_{10}(\Sigma_{\rm HI}) = A\times b + B$. Red dashed line with left axis: projected DM column density corresponding to the DM profile from \citet{Diemer_2023} for a halo with $M = 10^{12} \text{M}_{\odot}$ and $z = 1$.}
    \label{fig:col_density}
\end{figure}

\section{Covering fraction}
\label{sec:covering_fraction}

To further characterize the \MgII\ absorption, the covering fraction is derived for the 26 selected groups of more than five galaxies.
The covering fraction is commonly defined as the probability $P$ of detecting a \MgII\ absorption system at a given impact parameter from a galaxy or a group of galaxies. Practically, several methodologies are used in the literature to compute the covering fraction. \citet{Nielsen_2018} compute the covering fraction in impact parameters bins by doing the ratio of galaxies associated to an absorption by the total number of galaxies in that bin. \citet{Dutta_2020} use a cumulative covering fraction. \citet{Chen_2010} take into account how the gaseous halo scales with the B-band luminosity to normalize the impact parameter. Here, to be consistent with previous analysis performed on MEGAFLOW, we adopt the logistic regression method described in \citet{Schroetter_2021} to compute the differential covering fraction. This Bayesian method is particularly adapted in cases where bins would not be sufficiently or evenly populated. To describe it briefly, the probability $P$ of detecting a \MgII\ absorption system at a given impact parameter from a group is assumed to follow a logistic function of the form:

\begin{equation}
    P(\text{detection} = 1) \doteq L(t) =  \frac{1}{1+\exp{(-t)}},
	\label{eq:logistic}
\end{equation}
where $t$ is expressed as a function of the independent variables $X_i$ and of the model parameters $\theta$. In our case we consider that the variable is the impact parameter $b$ and that $t$ follows a logarithmic decrease of the form:

\begin{equation}
    t = f(X_i, \theta) = A(\log{b} - B).
	\label{eq:t}
\end{equation} 
The parameters of interest $A$ and $B$ are then fitted using a MCMC algorithm based on 9000 Bernoulli trials. This fit is performed using the pymc3 python module \citep{Hoffman_2011, Salvatier_2015}. Note that this method doesn't require any binning contrary to what can be found in other studies. In consequence our input are Booleans corresponding to the presence (or not) of an absorption.
In order to obtain a robust fit, two additional parameters are simultaneously fitted to take into account outliers: $f_{\rm out}$ is the fraction of outliers in the sample and $p_{\rm out}$ is the covering fraction associated to these outliers and assumed to be constant. The obtained best-fit parameters are listed in Table \ref{tab:cf}. 
\begin{center}
\begin{table}
\caption{Covering fraction fitted parameters for the two impact parameter definitions. The uncertainties are 2-$\sigma$.}
\label{tab:cf}
\begin{tabular}{|c c c c c|} 
 \hline
 & A & B & $f_{\rm out}$ & $p_{\rm out}$ \\ [0.5ex] 
 \hline
 $b_{\rm min}$ & -5.4$^{+4.3}_{-6.9}$ & 2.2$^{+0.6}_{-0.3}$ & 0.1$^{+0.4}_{-0.1}$ & 0.7$^{+0.3}_{-0.6}$ \\ 
 \\
 $b_{\rm center}/R_{\rm vir}$ & -2.7$^{+2.1}_{-6.8}$ & 1.7$^{+1.1}_{-0.9}$ & 0.2$^{+0.3}_{-0.2}$ & 0.6$^{+0.4}_{-0.6}$ \\ 
 \hline
\end{tabular}

\end{table}
\end{center}

\begin{figure}
  \includegraphics[width=8cm]{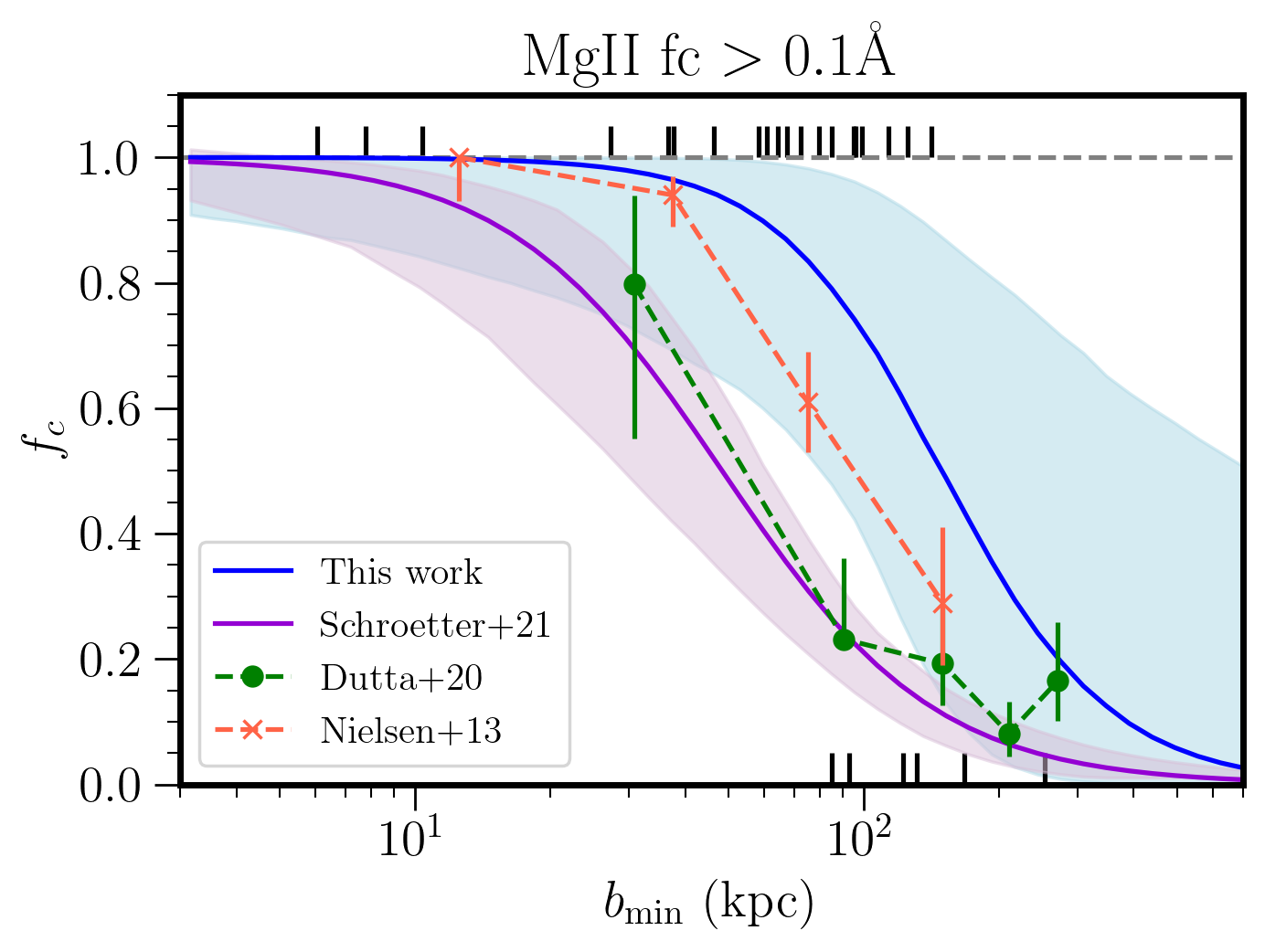}\hfill
  \includegraphics[width=8cm]{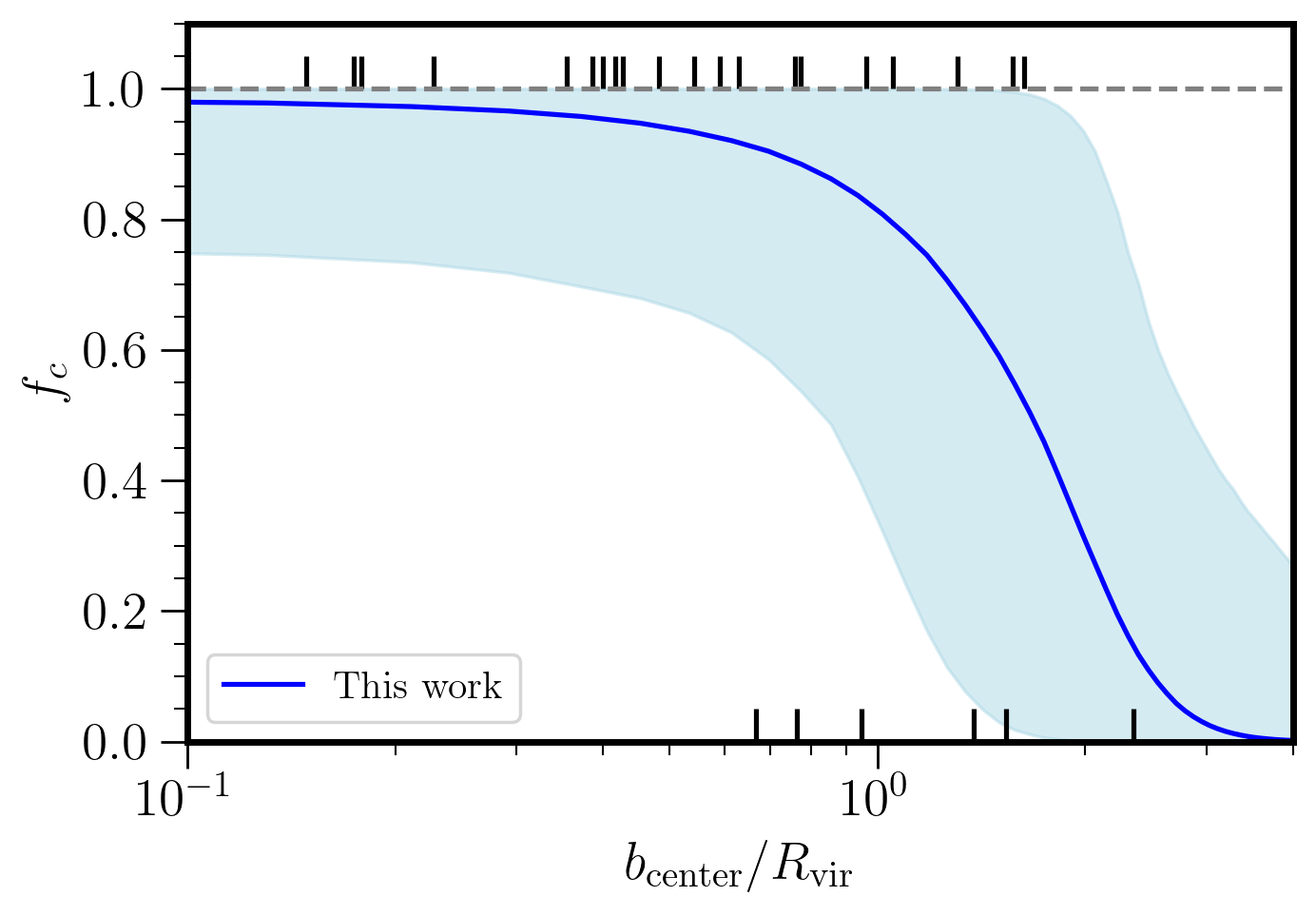}\hfill
    \caption{Differential covering fraction of \MgII\ absorption of width $W^{2796}_r > 0.1$\AA\ for the groups of five or more galaxies. Top: as a function of $b_{\rm min}$ and compared with the results from \citet{Schroetter_2021}, \citet{Dutta_2020} and \citet{Nielsen_2013}. Bottom: as a function of $b_{\rm center}/R_{\rm vir}$. Each vertical black mark corresponds to a group, it is equal to one if there is a counterpart absorption system and zero otherwise. The shaded areas correspond to the 95\% confidence level of the covering fraction. The error bars for \citet{Dutta_2020} and \citet{Nielsen_2013} correspond to the 68\% confidence level.}
  \label{fig:fc}
\end{figure}

We find that the $50\%$ covering fraction, namely $f_c(b)=0.50$, is reached for $\log_{10}(b_{\rm min}/\rm kpc) = 2.17 \pm 0.47$ (2-$\sigma$) and $b_{\rm center}/R_{\rm vir} = 1.67 \pm 0.98$ (2-$\sigma$). The fitted covering fractions as a function of $b_{\rm min}$ and $b_{\rm center}/R_{\rm vir}$ are plotted in Figure \ref{fig:fc} and are compared with the results from \citet{Schroetter_2021} and \citet{Nielsen_2013}.

\section{Discussion}
\label{sec:discussion}

As mentioned in Section \ref{sec:mgii_absorption}, three groups deviate significantly from the main $\REW - b_{\rm{center}}/R_{\rm{vir}}$ decreasing trend. Figure \ref{fig:group_catalog_3} gives us some hints on the particularities of these groups. The group 7 is below the relation. It's \MgII\ equivalent width is low in spite of being at small impact parameter from the LOS. This behaviour could be explained by the fact that four galaxies around the group center are quenched. The low star formation activity in the central part of this group is synonym of low galactic winds and, hence, low amount of gas ejected from the galaxies into the CGM. The group 18 is also below the main trend. It presents an elongated shape with five out of six galaxies aligned so that they could be part of a filament. In such case this group would not be virialized and the cool gas could then possibly be preferentially distributed along the filament. The group 28 at the contrary is above of the relation. It is a very compact group with small velocity dispersion leading to a low estimated virial mass. As it is composed of only five galaxies the uncertainty on the virial mass is large. In addition the group has no clear heaviest galaxy, so we estimated the position of the center as the barycenter of the group members. The position of the barycenter suffers from high uncertainties from the estimated stellar masses of the members. These combined uncertainties lead to a large error bar that could explain why this group is standing outside of the main relation.

Figure \ref{fig:group_catalog_3} also reveals very different kinds of group morphologies. For instance groups 8, 15, 20, 27, 28, 29, 30 are very compact both in projected and in velocity space meanwhile groups 14, 19 and 21 seem extended and diffuse. We also observe few groups with particularly elongated shapes like groups 12, 18, 33. These groups could be part of filaments accreting toward nodes of the cosmic web.

The absorption systems also present some diversity. In many cases like for groups 1, 4 and 6, all the components seem to be mixed and form a single absorption system with large velocity dispersion. In other cases such as 13, 24 and 28 we clearly observe distinct components, that are nonetheless difficult to attribute to a specific member. In few cases like groups 4, 18, 19 or 22 we can possibly identify the galaxy counterpart of some absorption components. For the group 19, we can clearly attribute a specific absorption component for four out of the five members. For the group 4 we can see in the spectra an absorption component matching with the galaxy 13, lying outside of the group (and that have been rejected by the halo refinement algorithm).

We also observe that for five groups out of 26, no counterpart \MgII\ absorption is found in the quasar spectra. For these five cases the estimated impact parameter to the center is relatively large which is consistent with the picture of a halo of cool gas vanishing at high distance.

\subsection{Comparison with field and isolated galaxies}
It is interesting to compare the covering fraction computed for our group sample to the covering fraction of field galaxies. For that we use the results from \citet{Schroetter_2021} that estimated the \MgII\ covering fraction for MEGAFLOW galaxies at redshifts $1< z < 1.5$ where both \MgII\ and \CIV\ absorptions could be observed with UVES. A total of 215 galaxies have been identified in this redshift range using their [\OII] emission. When multiple galaxies were present in the vicinity of an absorption system, they considered the impact parameter relatively to the closest galaxy. For that reason we compare their results to the covering fraction that we computed as a function of $b_{\rm min}$ (top panel of Figure \ref{fig:fc}). The fact that we use the same survey and the same methodology to compute the covering fraction allows a consistent comparison between our results. The overlap between our group sample and the sample used by \citet{Schroetter_2021} consists of five absorption systems out of the 52 that they used to compute their covering fraction. Finally we find that the covering fraction for groups is approximately three times larger than the one computed by \citet{Schroetter_2021} (the $50 \%$ covering fractions are reached respectively at 148 kpc versus 47 kpc).

\begin{figure}
	\includegraphics[width=7cm]{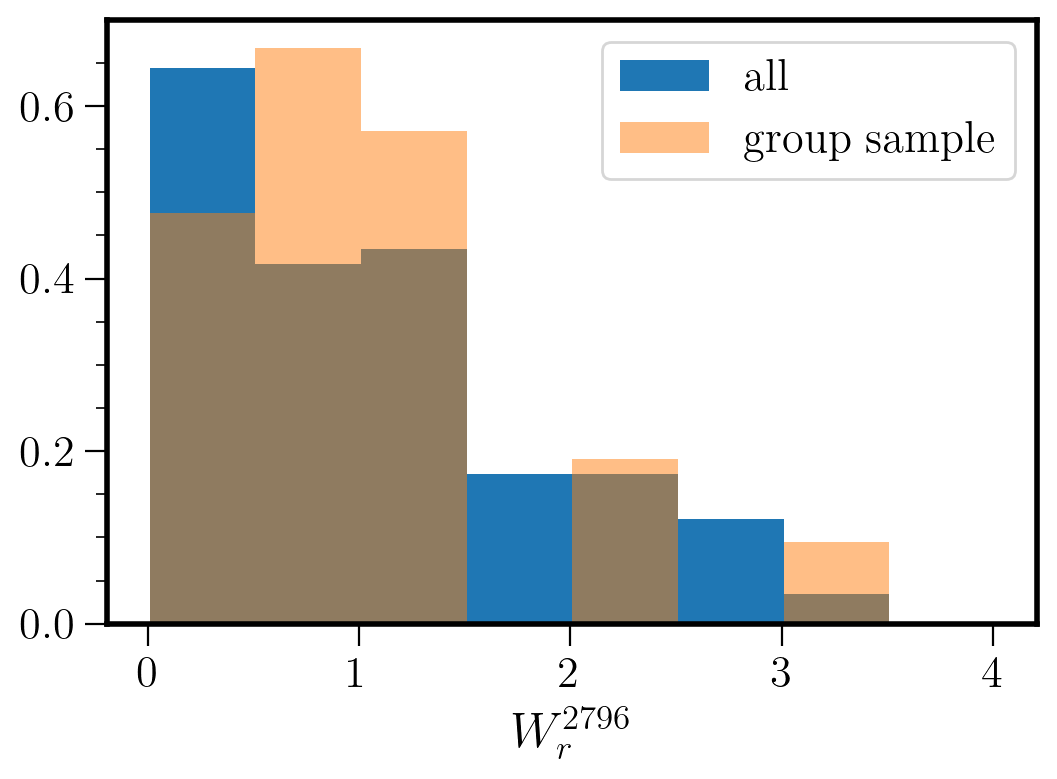}
    \caption{Distribution of \MgII\ absorption equivalent width for the 120 MEGAFLOW absorptions at $0.3 < z <1.5$ (in blue) and for the 21 groups of more than five galaxies (in orange) presenting absorption. The distributions have been normalized to be compared.}
    \label{fig:dNdW_megaflow}
\end{figure}

In terms of equivalent width, we observe that groups are not preferentially associated with strong absorptions in MEGAFLOW as shown in the \REW\ distribution presented in Figure \ref{fig:dNdW_megaflow}. Indeed, on the 59 strong absorptions with  $W^{2796}_r > 1.0$ \AA\ only nine are associated with groups of five galaxies or more. Reversely, on the six groups with an estimated virial mass above $10^{13} M_{\odot}$ only two present an associated absorption with $W^{2796}_r > 1$ \AA\ . Our results are in line with the works from \citet{Bouche_2006} and \citet{Lundgren_2009} that have shown that \REW does not grow with the mass of the halo but is rather anti-correlated with it.

We also compare our results to \citet{Dutta_2020}. In their section 3.5 they present the covering fraction computed for their full sample of 228 galaxies at redshift $0.8< z < 1.5$. There are two major differences with the work from \citet{Schroetter_2021}. First they did not select the quasar fields based on the presence of multiple \MgII\ absorptions as it has been done for MEGAFLOW, arguing that it would prevent their analysis from any bias due to pre-selection. Second, their sample is mostly composed of continuum-detected galaxies (it contains only 14 galaxies that have been identified from the research of emission lines in the vicinity of known \MgII\  absorptions). In their Figure 18 they show the covering fraction for their whole sample. When multiple galaxies are present around an absorption system they take into account all galaxies in their calculation. Their results show that the covering fraction is significantly affected by the choice of the absorption equivalent width limit. Nevertheless, in Figure \ref{fig:fc} we show that their covering fraction is completely consistent with the covering fraction computed by \citet{Schroetter_2021} on MEGAFLOW for an identical equivalent width limit of $0.1$ \AA.

It is also interesting to compare our result to the covering fraction estimated by \citet{Nielsen_2013} for isolated galaxies. They defined galaxies as isolated if they have no neighbours within a projected distance of 100 kpc and LOS velocity interval of $500~\kms$. They used 182 isolated galaxies at redshift $0.07< z < 1.12$ from the MAGIICAT sample which is built from a compilation of several galaxy-absorption pair samples (some of them consisting of galaxies identified around known \MgII\ absorption systems). They computed the covering fraction for several absorption equivalent width limits. In Figure \ref{fig:fc} we show their estimated covering fraction at $0.1$ \AA~. We observe that their covering fraction for isolated galaxies is significantly higher than the covering fraction obtained for field galaxies in the previously mentioned papers but remains lower than the results we find for groups, though within the 95\% confidence level.

\subsection{Comparison with literature about groups}

It is difficult to compare rigorously our results with the existing literature about groups, first because the definition of what is a group varies (for instance we do not consider pairs of galaxies as groups) and second because many different definitions/methods are used to estimate the covering fraction and could have impacts on the results. Nonetheless we can perform a qualitative comparison. \citet{Nielsen_2018} studied the groups in the MAGIICAT sample. They show that the overall covering fraction (without taking into account the effect of the impact parameter) is higher for groups (at the 2.2 $\sigma$ level) than for isolated galaxies. They also show that the \MgII\ equivalent width is consistent with the superposition model proposed by \citet{Bordoloi_2011} but that the absorption kinematics reveal a more complex behavior and make them favor the hypothesis that the absorptions are caused by an intragroup medium rather than by individual galaxies. This assumption is consistent with our finding that the extent of the \MgII\ halo seems to scale with the mass (hence the virial radius) of the halo.

\citet{Dutta_2020} and \citet{Dutta_2021} also studied the impact of environment on the \MgII\ covering fraction at z $\approx$ 1. They find that the covering fraction around groups is three times higher than around isolated galaxies. This result is in line with our conclusion even if their definition of what is a group and their way to compute the covering fraction is different.

Finally, the interpretation of our results on groups along with the existing literature lead to the following picture:
\begin{itemize}
    \item Absorptions are mostly caused by individual or small ensemble of four or less galaxies compatible with their natural correlation in the field. In MEGAFLOW only 21 out of 120 $z<1.5$ absorptions are caused by groups of more than five galaxies.
    \item The \REW\ of absorptions associated with over-densities are not higher (Figure \ref{fig:dNdW_megaflow}). This is consistent with the results from \citet{Bouche_2006}, \citet{Lundgren_2009}, \citet{Gauthier_2009}, that rather find an anti-correlation with the halo mass. Strong absorptions would hence be preferentially caused by un-virilized clouds of gas mostly due to strong outflows around starburst galaxies. At the contrary the quenching of galaxies as they enter groups lead to less extreme galactic winds and more virialized clouds.
    \item However, the spatial extent of \MgII\ is higher for more massive halos, as \REW drops at the virial radius.
    \item The probability to find an absorption is much higher for dense environments (21 groups out of 26 are associated with an absorption, meanwhile the 101 remaining absorptions of MEGAFLOW are distributed between more than $\approx 1000$ galaxies).
\end{itemize}

\subsection{Potential effect of the quasar field pre-selection}
One could object that the pre-selection of quasar line-of-sights based on the presence of multiple strong absorptions (\REW\ $> 0.5$ \AA~) could introduce a bias in the measurement of the covering fraction presented here. We believe that if it exists, this bias is small for the following reasons.

First, if a bias were present in MEGAFLOW it would have been seen in the analysis of \citet{Schroetter_2021} for field galaxies. However, the covering fractions computed by \citet{Schroetter_2021}  and the covering fraction from \citet{Dutta_2020} (on randomly selected LOS) or \citet{Lan_2020} are all very similar.

Second, as shown in \citet{Schroetter_2021}, the \MgII\ equivalent width distribution $(\mathrm{d}n/\mathrm{d}W)$ in MEGAFLOW follows the same exponential law ($\propto \exp(-W_{r}/W_{0})$) as found in random sight-lines \citep[e.g][]{Nestor_2005, Zhu_Menard_2013} but with a boosted normalization. Hence, even if there were a relation between the galaxy properties and the \MgII\ absorption equivalent widths, the MEGAFLOW pre-selection procedure doesn't introduce a bias in the covering fraction.

Third, the covering fraction we compute for groups covers a very wide redshift range (0.3-1.5) with $\approx$ 4000 spectral channels, or $\approx$ 2000 independent possible redshifts given the MUSE resolution. The MEGAFLOW survey has $\approx$ 100 galaxies per field, of which $\approx$ 50-60 are at these low redshifts. Hence, having 3, 4 or 5 pre-selected absorptions might be affecting the covering fractions of 5-10\% of the samples.
In other words, there are no reasons to presume a strong bias due to the absorption pre-selection.

Finally, we performed a quantitative experiment using a simple toy model presented in appendix \ref{appendix:toy_model} to mimic the effect of the line-of-sight pre-selection based on the presence of multiple strong absorptions. For a sample of $\approx 20$ selected fields (similar to what we have in MEGAFLOW) populated by $\approx 60$ galaxies each, we only observe a small shift in the measured covering fraction, compatible with the 2-$\sigma$ measurement error. With a sample ten times larger, this shift is significant at the 3.3-$\sigma$ level. Finally we conclude that if existing, the bias would be at most 5-10$\%$ which is small compared to the factor three that we observe between the covering fraction of groups versus field galaxies.

Bouché et al. 2023 (in prep) present an alternative model to estimate the effect of sight-lines pre-selection. They find that the field pre-selection has negligible effects on the measured covering fraction. They also reproduce the distribution of \MgII\ absorption equivalent widths ($\mathrm{d}N/\mathrm{d}W$) and show that it is not affected by the selection process.

\subsection{limitations and future prospects}
The work presented here has several limitations. 
The first one is that, as can be seen from Figure \ref{fig:group_catalog_3} some groups are probably cropped by the FOV. In such cases the group center that we identified could be wrong, as well as the impact parameter relative to the quasar LOS. The impact of this effect is difficult to quantify and has not been taken into account in this work. However the fact that our group centers often match with one or several passive galaxies (as observed in the literature) makes us confident about the robustness of our group finding procedure.

The second one is that the redshift dependency of our results has not been investigated given the size of our sample. A possible improvement would be to increase the statistics and to fit the covering fraction as a function of both the impact parameter and redshift.

This work is focused on groups of 5 or more galaxies. We justified this choice by the analysis of the two point correlation function  that reveals that the typical number of galaxies expected around an absorption system is $\approx 3$ for the MUSE FOV. As we wanted to study over-densities, we focused on groups with a number of galaxies higher than this value. In addition, we wanted to derive the mass of the groups using the velocity dispersion of the galaxies. That method requires a sufficient number of galaxies. However, our FoF algorithm finds 93 groups having 3 to 5 galaxies. An extension of this work could be to investigate in more detail the absorptions in quasar sight-lines in the vicinity of these smaller groups.

Finally, a detailed case by case analysis of the identified group-absorption pairs taking advantage of the UVES high resolution spectra would be interesting and is planned to be explored in a future paper.

\section{Conclusions}

We presented our results about the cool gas traced by \MgII\ around groups of galaxies in the MEGAFLOW survey. MEGAFLOW is based on observations from VLT/MUSE and VLT/UVES of 22 quasar fields presenting multiple ($\geq 3$) strong \MgII\ absorptions. A total of 1208 galaxies were detected in the foreground of quasars, both from their continuum and emission lines (mainly [OII]), with estimated $\log_{10}(M^*/\text{M}_{\odot})$ ranging from 6 to 12 and redshift ranging from 0.1 to 1.5.

Using a combination of a FoF algorithm and a halo occupation algorithm we identified a total of 33 groups of more than 5 galaxies. Among them 26 are located at the foreground of the quasars and can be used to study counterpart \MgII\ absorptions within quasar spectra. These groups have $10.8 < \log_{10}(M/\text{M}_{\odot})< 13.7$ and $0.4 < z< 1.5$. The analysis of the group properties and their counterpart \MgII\ absorptions led to the following conclusions:

\begin{enumerate}
    \item On the 120 \MgII\ absorption systems present in MEGAFLOW at $z<1.5$, 21 could be associated with a group of more than five galaxies.
    \item For five groups of more than five galaxies, no \MgII\ absorption has been detected in the nearby quasar spectrum down to a detection limit of \REW\ $\approx 0.1$ \AA.
    \item The $W^{2796}_r$ appears to be clearly anti-correlated with the impact parameter. It drops at $\approx 150$ kpc from the closest galaxy and $\approx R_{\rm vir}$ suggesting that \MgII\ halos scale with halo masses.
    \item The \MgII\ covering fraction measured for groups is $\approx 3$ times higher than the one computed for field galaxies. This result is consistent with other recent literature results.
    \item However contrary to some other studies we do not find that $W^{2796}_r$ is higher in groups. It suggests that strong absorptions are preferentially caused by outflows induced by individual star forming galaxies rather than by accumulation of gas in the intragroup medium.
    \item We derived \HI\ column densities from $W^{2796}_r$ and compared them to the dark matter column density profile for a halo of similar mass. The \HI\ and DM profiles exhibit a very similar shape with a clear drop at the virial radius.
    \item The groups present various morphologies: compact, diffuse, filamentary or irregular. The associated absorption systems are also diverse. They contain multiple absorption components that are difficult to attribute to individual galaxies.
\end{enumerate}

\label{sec:conclusion}




\section*{Acknowledgements}
This work has been carried out thanks to the support of the ANR 3DGasFlows (ANR-17-CE31-0017).

The calculations and figures have been made using the open-source softwares NUMPY \citep{Numpy_2011, Numpy}, SCIPY \citep{Scipy}, MATPLOTLIB \citep{Matplotlib}, ASTROPY \citep[][]{Astropy2013, Astropy2018} and PYMC3 \citep[][]{Hoffman_2011, Salvatier_2015}.

The data used in this work are based on observations made with ESO telescopes at the La Silla Paranal Observatory.

\section*{Data Availability}
The underlying data used for this article are available in the ESO archive (\url{http://archive.eso.org}).\\
The MEGAFLOW catalog will be soon available at the address \url{https://megaflow.univ-lyon1.fr/} and be visualized at \url{https://amused.univ-lyon1.fr}.\\
The data generated for this work will be shared on reasonable request to the corresponding author.



\bibliographystyle{mnras}
\bibliography{example} 




\appendix

\section{Groups visualisation}

The 33 groups with more than 5 galaxies are presented in Figure \ref{fig:group_catalog_3}. The left column shows the galaxies in projected space. The middle column shows the galaxies in phase space. The right column shows the UVES spectra of the corresponding quasars at the redshift of the groups when available. Passive galaxies (with $sSFR < 0.1$ Gyr$^{-1}$) are colored in red.

\begin{figure*}
	\includegraphics[width=14cm]{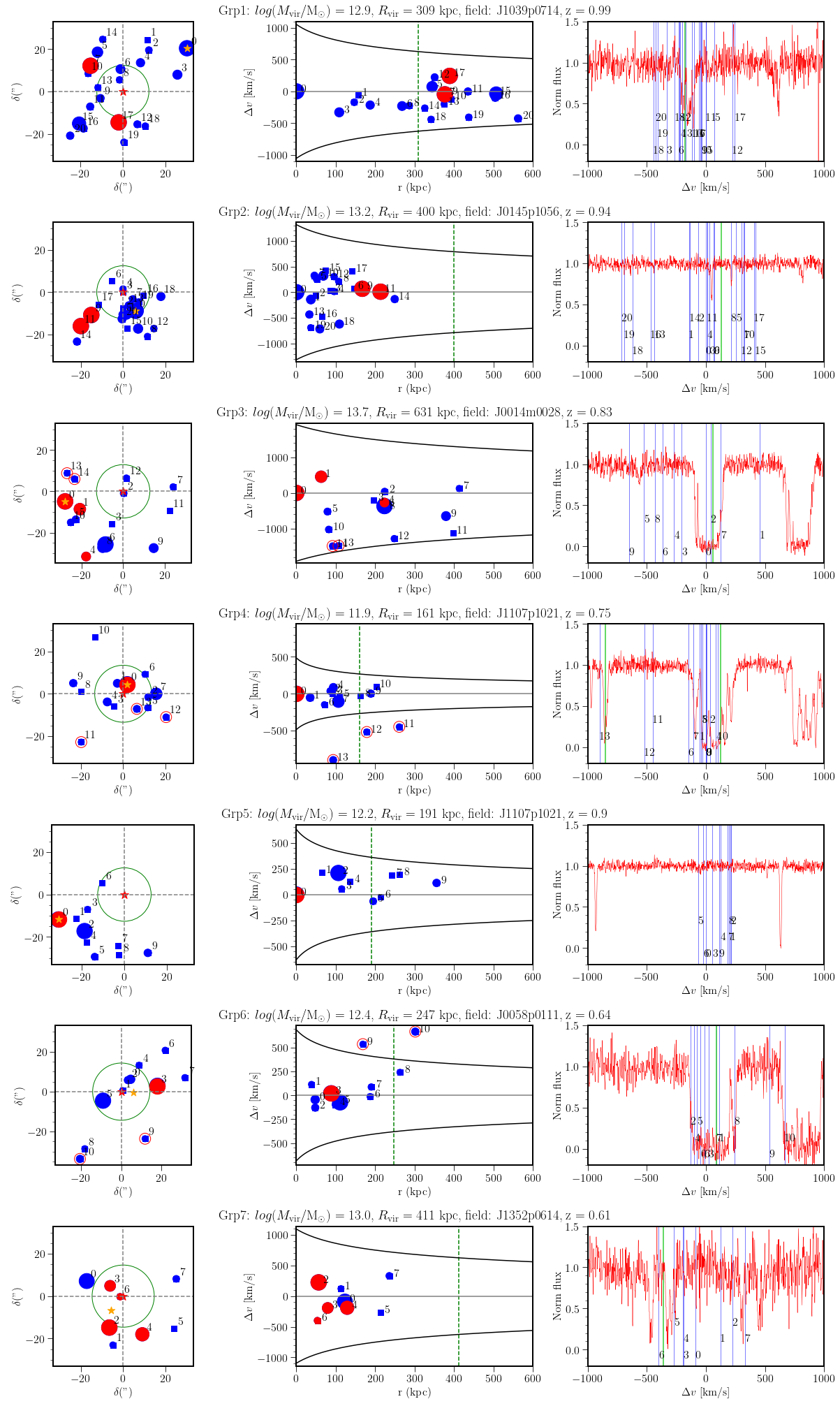}
    \label{fig:group_catalog_1}
\end{figure*}

\begin{figure*}
	\includegraphics[width=14cm]{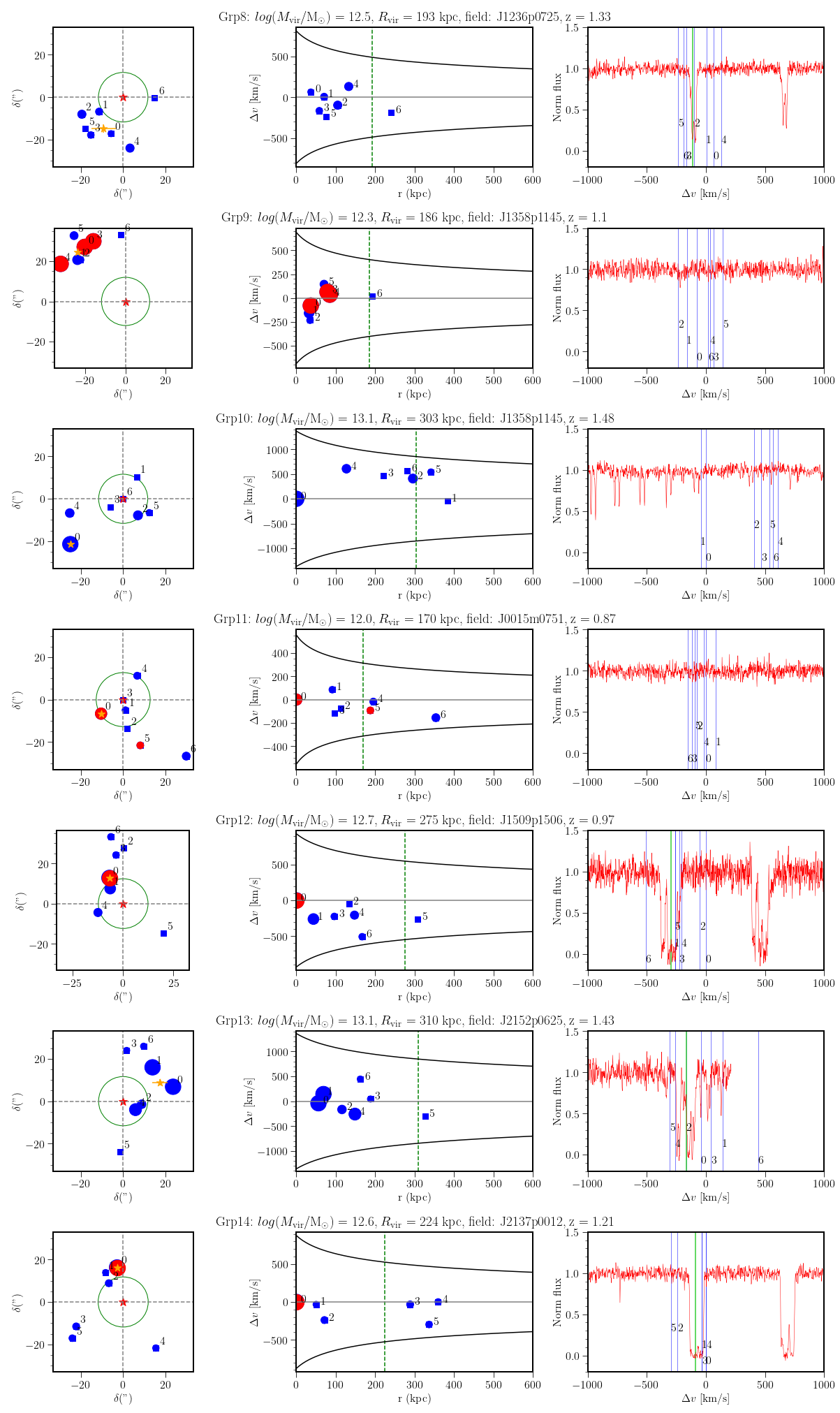}
    \label{fig:group_catalog_2}
\end{figure*}

\begin{figure*}
	\includegraphics[width=14cm]{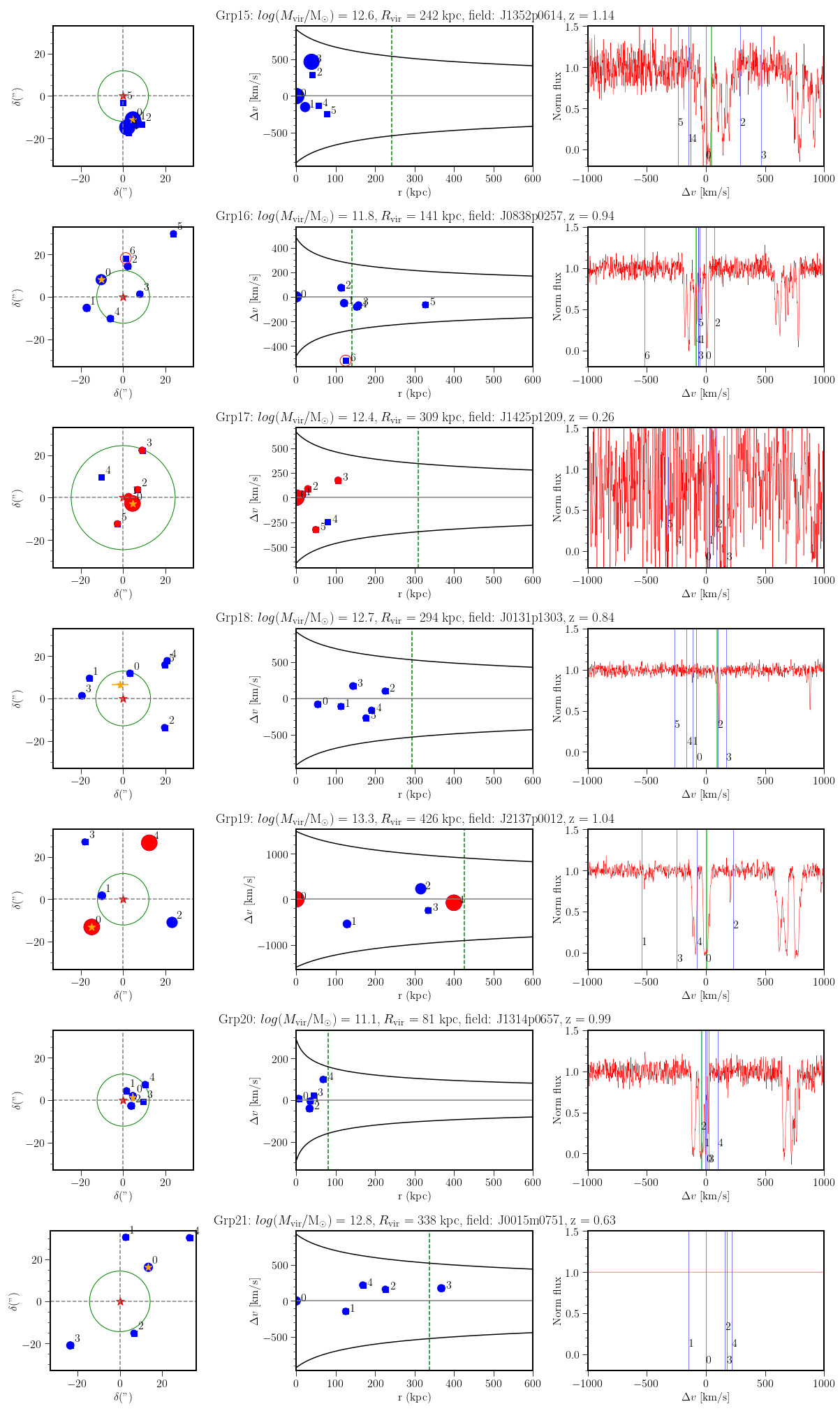}
    \label{fig:group_catalog_3}
\end{figure*}

\begin{figure*}
	\includegraphics[width=14cm]{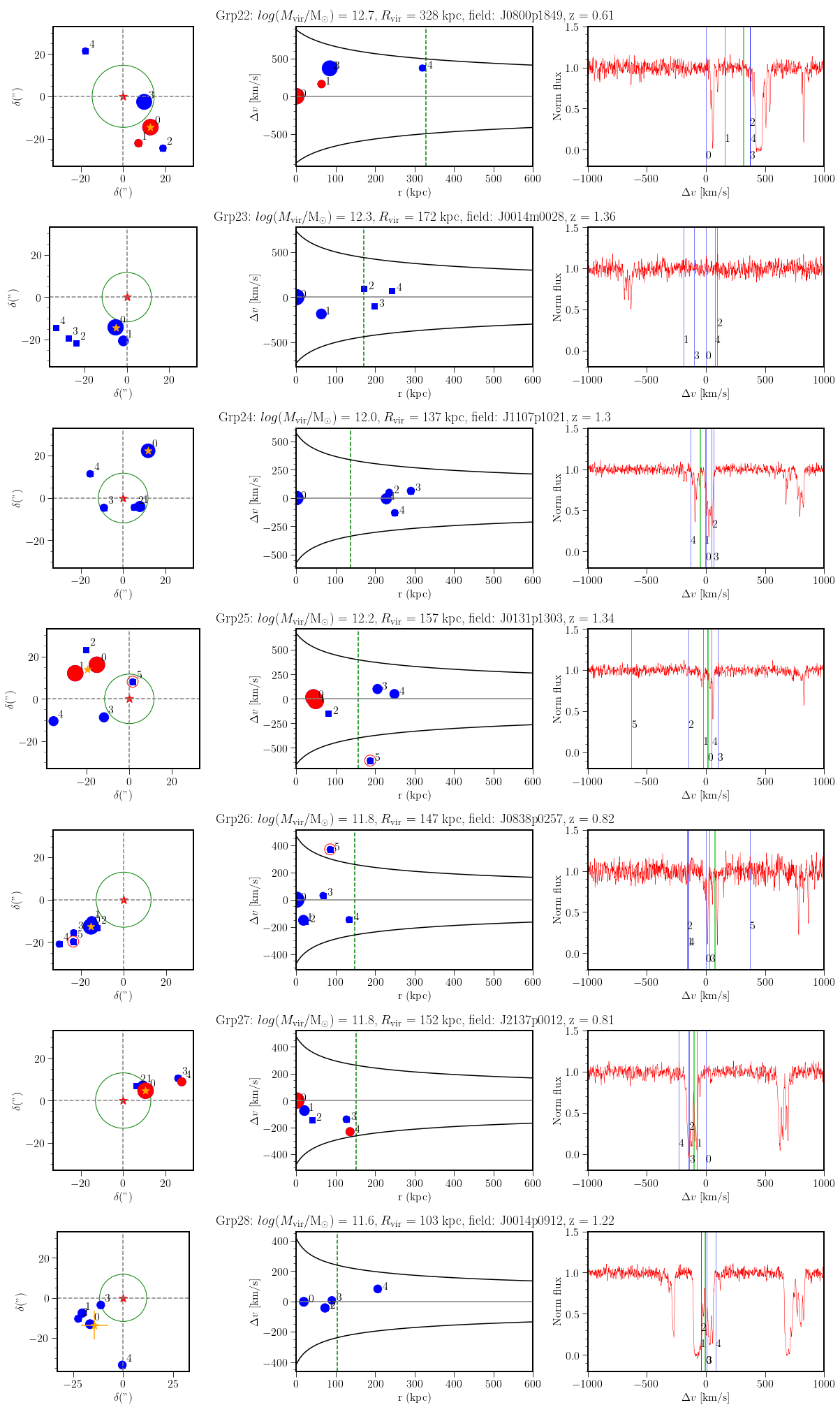}
    \label{fig:group_catalog_4}
\end{figure*}

\begin{figure*}
	\includegraphics[width=14cm]{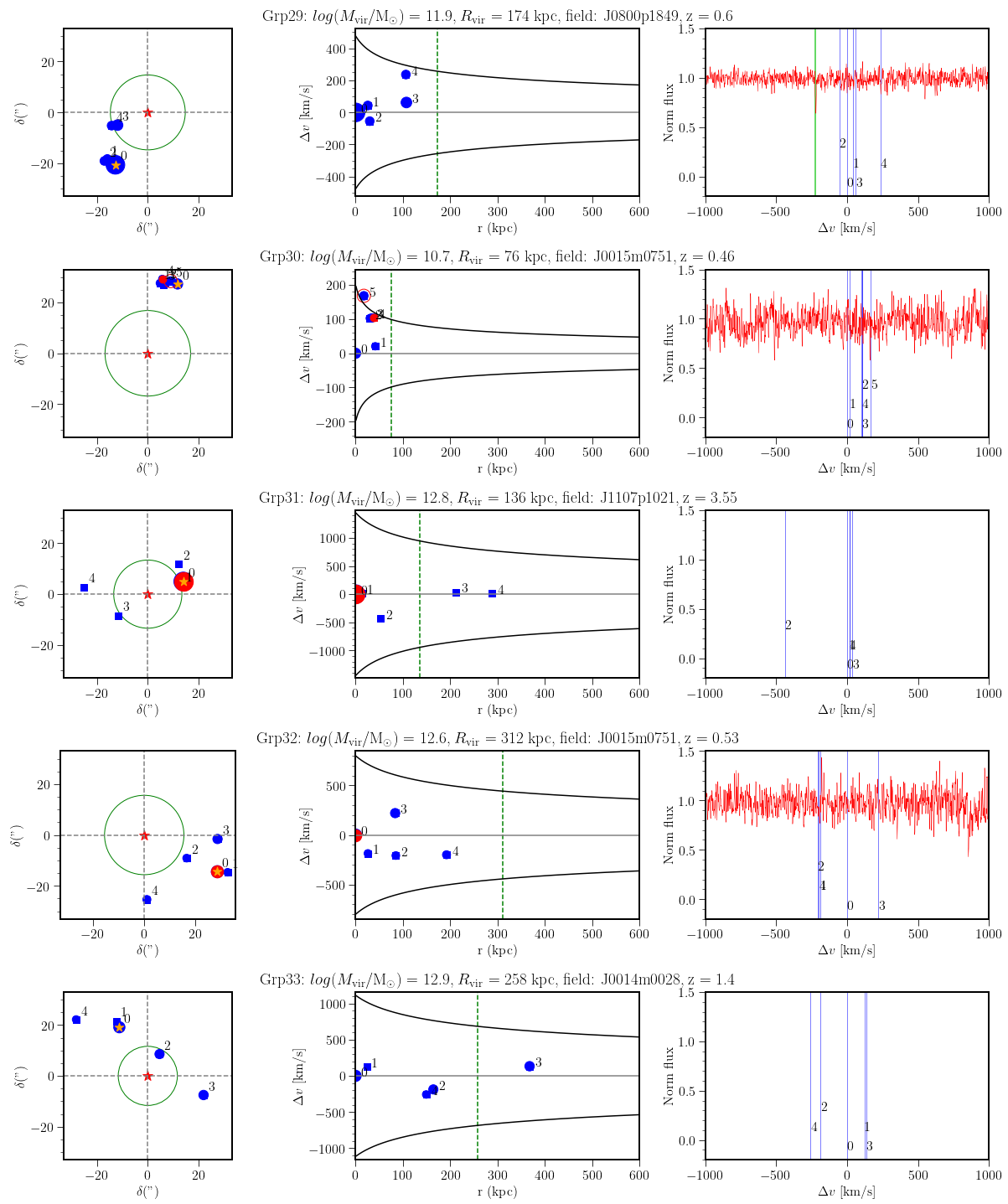}
    \caption{Visualization of the individual groups. Left column: groups in projected coordinates (right ascension and declination). The dots are the galaxies, with a size proportional to the log of their estimated stellar mass. The red dots are the "passive" galaxies with a sSFR $< 0.1$~Gyr$^{-1}$. The galaxies circled in red are the galaxies that have been excluded from the group by the halo occupation method. The orange cross is the group center. The red star at (0,0) is the quasar. The green circle represents a 100 kpc radius around the quasar. Middle: the galaxy distribution in phase space (distance to the center of the group along the x-axis and velocity separation to the center of the group along the y-axis). The dashed vertical line is the estimated virial radius. The black lines are the escape velocity caustics computed from the estimated mass of the groups assuming NFW properties. Right: high-resolution spectra of the central quasar. The x-axis represents the velocity difference relative to the center of the group. The green vertical line is the estimated \MgII\ absorption velocity difference. The blue lines are the velocity differences of the galaxies in the group.}
    \label{fig:group_catalog_5}
\end{figure*}


\section{About the effect of the MEGAFLOW selection}
\label{appendix:toy_model}

As discussed in Section \ref{sec:megaflow},
the MEGAFLOW survey is built around a sample of quasar
sight-lines with multiple ($N=3,4,5$) \MgII{} systems at $0.3<z<1.5$ \citep[][Bouché et al., in prep.]{Schroetter_2019, Zabl_2019}.
This pre-selection of sight-lines might introduce a bias in the measurement of the covering fraction.
In \S~7.4, we present qualitative arguments against such bias. Here, we quantify the potential bias on the covering fraction caused by the pre-selection of quasar sight-lines, which is only relevant for galaxy-based analysis such as the covering fraction in galaxies \citep{Schroetter_2021} or groups, as in this paper.

To quantify such bias, the idea  is to build a toy model that mimic a sample of MUSE fields randomly populated with galaxies and to select those with more than 3 strong absorptions. We then estimate the covering fraction for these selected fields and compare it to the one that we would have without selection.

\begin{figure}
	\includegraphics[width=1.0\columnwidth]{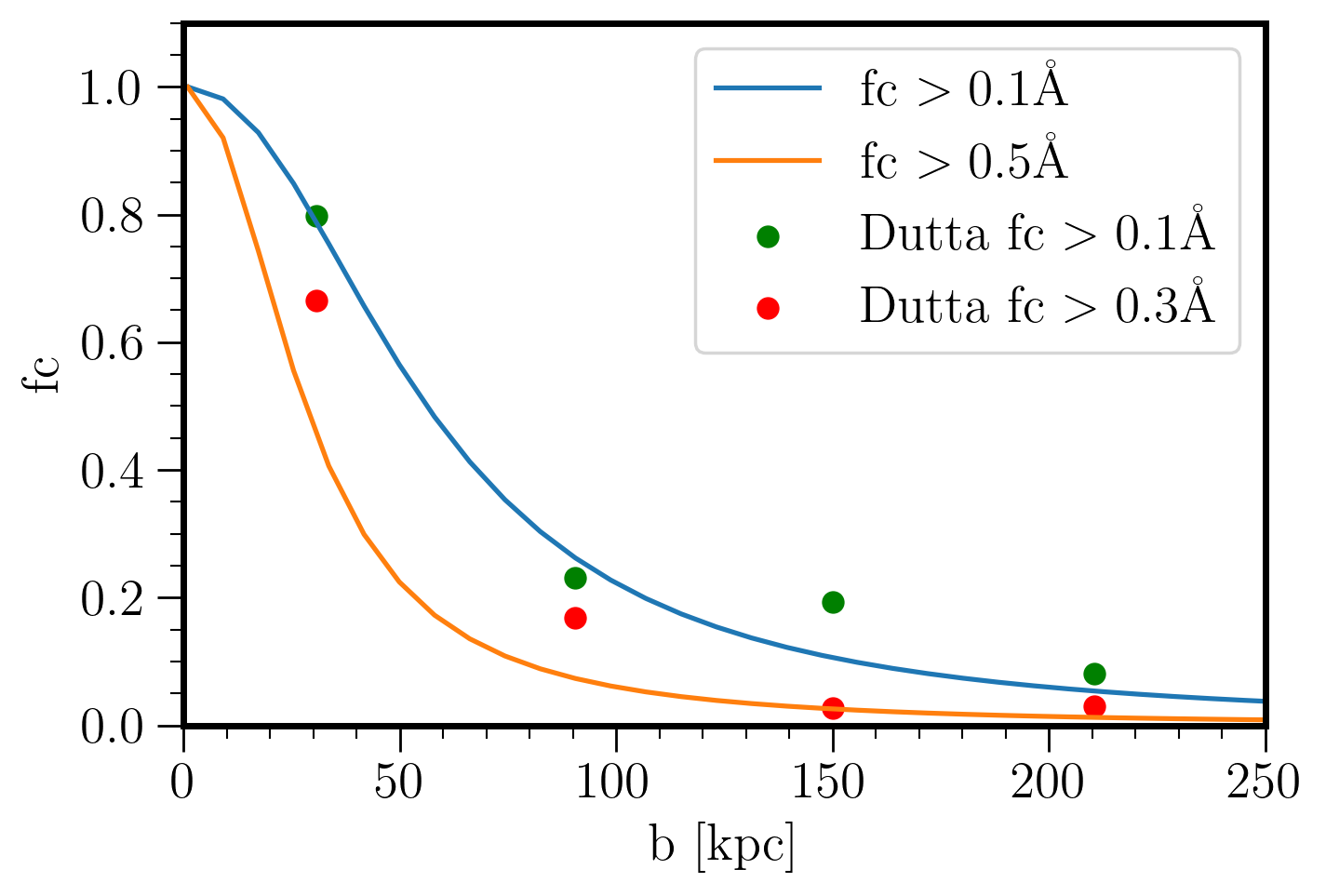}
    \caption{Assumed universal differential covering fractions used for the toy model for 0.1\AA\ (orange) and 0.5\AA\ (blue) detection limits. These assumed covering fraction are consistent with the differential covering fraction presented by \citet{Dutta_2020}.}
    \label{fig:toy_fc_assumption}
\end{figure}

Specifically, we considered 50 fields of view of $500\times 500$ kpc (similar to the MUSE FOV at $z \approx 1$) that we assumed centered on a quasar sight-line. We then populate each of these fields with $N$  galaxies (from a Poisson distribution of parameter $\lambda=60$) with random projected coordinates. For each galaxy we then assign two Boolean flags  representing respectively the presence of a weak (>0.1\AA) and a strong (>0.5\AA) counterpart \MgII\ absorption. For that we use two different covering fractions (shown in Figure \ref{fig:toy_fc_assumption})  based on the results from \citet{Dutta_2020}.

We then select the fields with at least 3 strong absorptions (with $\REW>0.5$~\AA) to mimic the MEGAFLOW pre-selection. There are 21 such fields, which is similar to the number of fields in MEGAFLOW.

\begin{figure}
	\includegraphics[width=1.0\columnwidth]{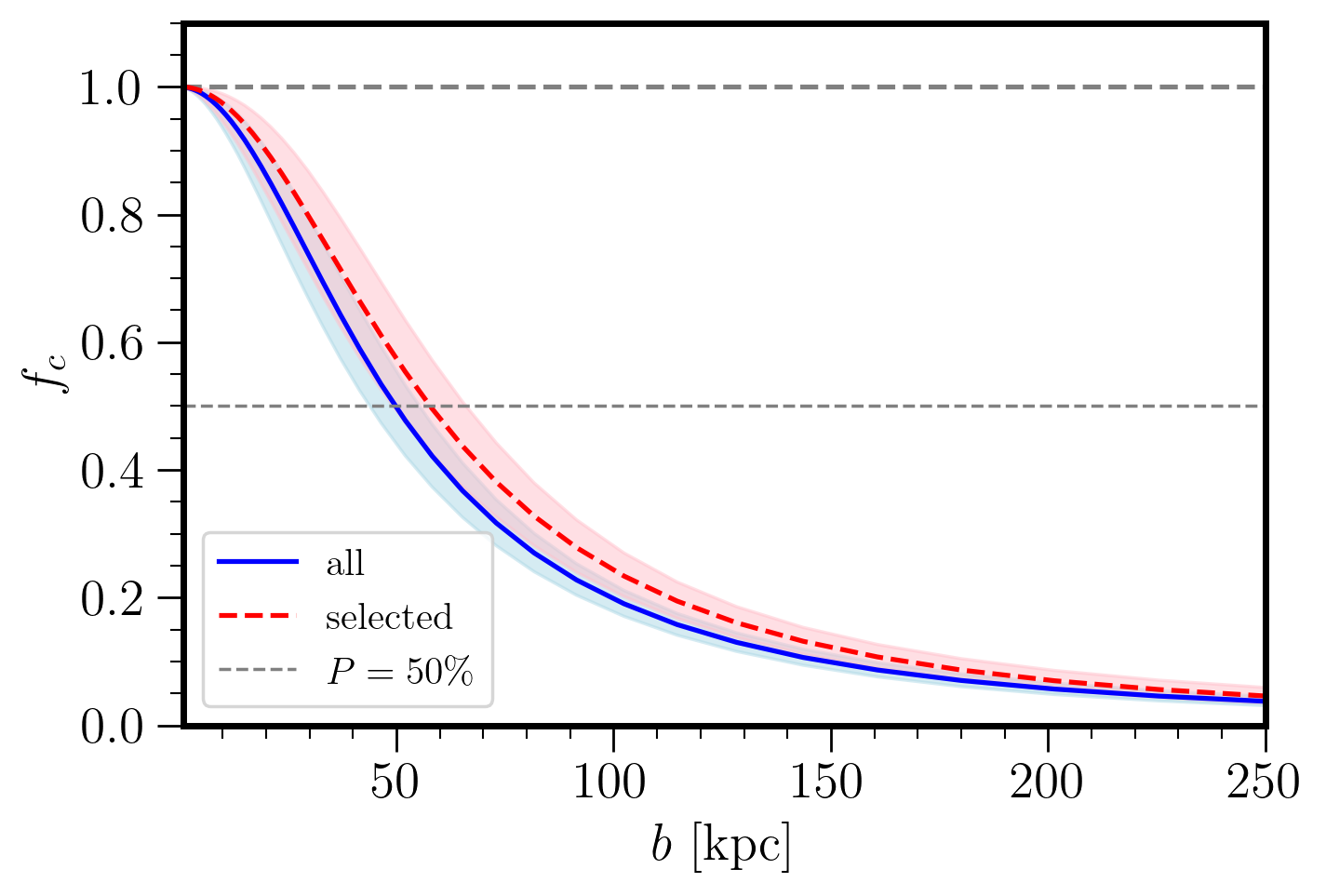}
    \caption{Comparison of the computed 0.1\AA\ covering fraction for the selected sample versus for the whole sample.}
    \label{fig:toy_fc}
\end{figure}

Finally, in these selected fields, we recompute the 0.1~\AA\ covering fraction and  compare it to the 0.1~\AA\ covering fraction for the whole sample. To compute the 0.1~\AA\ covering fraction, we use the method described in \citet{Schroetter_2021} in Section \ref{sec:covering_fraction}. Figure \ref{fig:toy_fc} shows the covering fraction on the pre-selected sample compared to the one on the whole sample with the computed 2-$\sigma$ intervals. Finally we find that the $0.1$ \AA~ covering fractions reach $50 \%$ at $57.8^{+8.2}_{-8.6}$ kpc and $49.8^{+5.5}_{-5.9}$ kpc (the indicated error are 2-$\sigma$) respectively for the selected and the whole sample. The two are statistically compatible with each other within the 2-$\sigma$ uncertainties. In order to have more statistics, we perform the same experiment with 500 fields (that yields $\approx$ 200 pre-selected fields). We find that the $0.1$ \AA~covering fractions reach $50 \%$ at $61.4^{+2.5}_{-2.5}$ kpc and $56.3^{+1.8}_{-1.8}$ kpc respectively for the selected and the whole sample. The two are values are in tension at the 3.3-$\sigma$ level. So finally we conclude that the pre-selection might introduce a small bias of 5-10\%. This bias is negligible compared to the factor three that we observe between the covering fraction of groups versus field galaxies.

Some tests on the parameters of this toy model reveal that the magnitude of this relatively low significance shift is mainly driven by the slope of the covering fraction. Indeed, the slope of the covering fraction encodes how the gas halos are similar with each others. The steeper is the slope, the smaller are the effects of the field selection because the presence of \MgII\ is hence mostly determined by the impact parameter and doesn't depends on galaxy properties.

\bsp	
\label{lastpage}
\end{document}